\UseRawInputEncoding
\documentclass[twocolumn]{aastex62}

\submitjournal{ApJ}
\usepackage{amsmath}
\usepackage{comment}
\usepackage{cleveref}

\crefformat{section}{\S#2#1#3} 
\crefformat{subsection}{\S#2#1#3}
\crefformat{subsubsection}{\S#2#1#3}

\newcommand{\lsim}{\raisebox{-0.13cm}{~\shortstack{$<$ \\[-0.07cm] $\sim$}}~}
\newcommand{\gsim}{\raisebox{-0.13cm}{~\shortstack{$>$ \\[-0.07cm] $\sim$}}~}

\shortauthors{Caputi et al.}

\begin{document}

\title{\bf {Pseudo Little Red Dot: an Active Black Hole Embedded in a Dense and Dusty,  Metal-Poor Starburst Galaxy at $\lowercase{z}=5.96$}}

\shorttitle{Pseudo-LRD: an Active BH in a Dense and Dusty, Metal-Poor Starburst Galaxy at $\lowercase{z}=5.96$}

\correspondingauthor{Karina I. Caputi}
\email{karina@astro.rug.nl}

\author[0000-0001-8183-1460]{Karina I. Caputi}
\affiliation{Kapteyn Astronomical Institute, University of Groningen,
P.O. Box 800, 9700AV Groningen,
The Netherlands
}

\author[0009-0000-0413-5699]{Ryan A. Cooper}
\affiliation{Kapteyn Astronomical Institute, University of Groningen,
P.O. Box 800, 9700AV Groningen,
The Netherlands
}

\author[0000-0002-5104-8245]{Pierluigi Rinaldi}
\affiliation{Space Telescope Science Institute, 3700 San Martin Drive, Baltimore, MD 21218, USA}

\author[0000-0001-6066-4624]{Rafael Navarro-Carrera}
\affiliation{Kapteyn Astronomical Institute, University of Groningen,
P.O. Box 800, 9700AV Groningen,
The Netherlands
}

\author[0000-0001-8386-3546]{Edoardo Iani}
\affiliation{Institute of Science and Technology Austria (ISTA), Am Campus 1, 3400 Klosterneuburg, Austria}

\author[0009-0008-3005-0435]{Abigail A. Tumborang}
\affiliation{Kapteyn Astronomical Institute, University of Groningen,
P.O. Box 800, 9700AV Groningen,
The Netherlands
}

\begin{abstract}

We present a study of {\it Pseudo-LRD-NOM} ({\it Pseudo little red dot with no metal lines}), a highly magnified low-mass galaxy behind the lensing cluster Abell 370  at $\lowercase{z}=5.96$. We classify this object as a pseudo-LRD because its red rest-frame optical colour is contaminated by a prominent H$\alpha$ line  (with $\rm EW_0 \gsim 800 \, \AA$) present in its JWST NIRSpec spectrum. H$\alpha$ is dominated by a narrow component and also has a minor broad component indicative of an active black hole with $M_{\rm BH} \approx 5.6 \times 10^6  \,M_\odot$. A narrow $\rm H\beta$ emission line is also detected (with $\rm S/N = 8$), producing a Balmer decrement (narrow) $\rm H\alpha/H\beta = 11$. The rest-frame UV spectral slope is $\beta_{\rm UV}^{\rm spec}=-1.20 \pm 0.28$.  All these features can be ascribed to high dust attenuation. However,  no [OIII]$\lambda5007$ or any other metal lines are detected in the spectrum, so [OIII]5007/$\rm H\beta < 0.25$, at odds with a simple dust-attenuation explanation. Accounting for all the spectral properties requires the model of a starburst with moderate colour excess E(B-V)$\approx0.2-0.5$, high gas density ($n_{\rm H} \gsim 10^6 \, \rm cm^{-3}$) and low to extremely low gas/stellar metallicities ($Z = 0.01-0.1 \, \rm Z_\odot$). The demagnified stellar mass is $2.25^{+1.30}_{-0.83} \times 10^7 \, \rm M_\odot$  and the stellar-mass surface density is $\Sigma_\ast = 418^{+725}_{-310} \, \rm M_\odot / pc^2$, similar to that of massive/nuclear star clusters. {\it Pseudo-LRD-NOM} provides evidence of massive black-hole growth occurring in a high-density, dusty starburst which is at the early stages of its chemical enrichment, and is likely a precursor to a real LRD.

\end{abstract}


\section{Introduction}
\label{sec:intro}

Understanding the formation of the galaxies that we see in the Universe today requires tracing their building blocks in cosmic time. These smaller units carried the imprint of the first stellar populations formed in those galaxies, as well as of the initial steps of chemical enrichment \citep[e.g.,][]{Leboutellier2013}. At the same time, they are supposed to be the sites where black holes formed and efficiently started to grow \citep{greene_intermediate-mass_2020}.  Therefore, spectroscopic studies of low stellar-mass galaxies (with $M_\ast< 10^9 \, \rm M_\odot$) at high redshifts can provide the key to unveiling the very basis of galaxy evolution.

The spectral properties of low stellar-mass galaxies, commonly called \textit{dwarf galaxies} due to their relatively small sizes, have been extensively studied at low redshifts \citep[e.g.,][]{Oestlin2001,Guseva2009,PMontero2013,Amorin2014,Yang2017}. They are typically characterised by prominent emission lines, indicative of high ionization states at sub-solar metallicities. In the rare cases that the metallicity is very low \citep[e.g.,][]{Izotov2024},  the metal lines are weak or absent, resembling the conditions that are expected to have been more common at early cosmic times.

\begin{figure*}[ht!]
    \centering
    \includegraphics[width = 0.98 \textwidth]{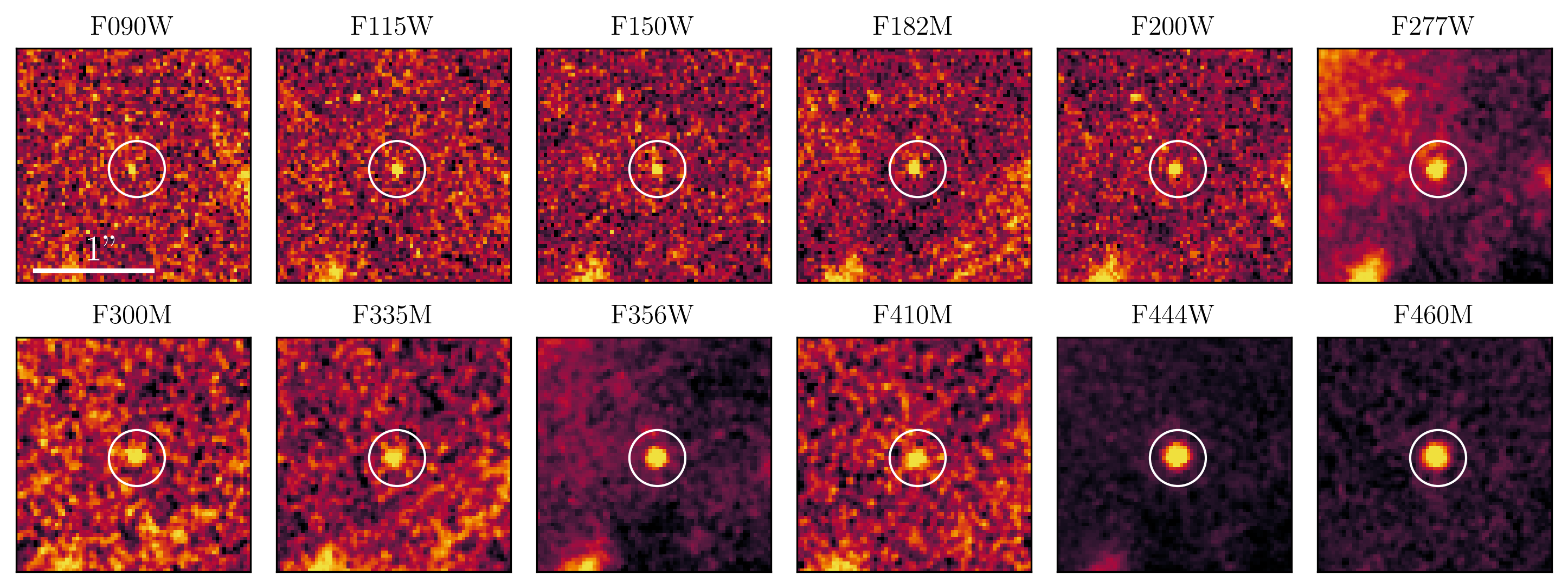}
    \caption{Postage stamps of {\it Pseudo-LRD-NOM} in different JWST/NIRCam bands. The two reddest filters shown, i.e., F444W and F460M, encompass the bright H$\alpha$ line.} 
    \label{fig:stamps}
\end{figure*}

However, until recently, spectroscopic observations of low-mass galaxies at high $z$ were very challenging for existing telescopes, and thus only a handful of cases behind lensing fields could be studied \citep[e.g.,][]{Vanzella2016, Karman2017,Vanzella2021}. Thanks to the advent of JWST \citep{gardner_james_2023}, the study of low-mass galaxies up to the Epoch of Reionization has now become routinely possible even in non-lensed fields \citep[e.g.,][]{Schaerer2022, Castellano2024, endsley_star-forming_2024,Harshan2024,iani_midis_2024-1,Rinaldi2025_emergence,Trussler2025}. 

In parallel, another high-$z$ galaxy population whose study has boomed in the JWST era is that of the so-called Little Red Dots \citep[e.g.,][]{kokorev_census_2024,matthee_little_2024}. These sources are characterised by being very compact \citep[e.g.,][]{Baggen2024} and having a V-shaped $f_\lambda (\lambda)$ spectral energy distribution \citep[SED; e.g.,][]{Kocevski2025}. Virtually all of them harbour broad-line active galactic nuclei \citep[BLAGN; e.g.,][]{greene_uncover_2024,Hviding2025}.  Based on the Balmer-line broadening, the central black holes appear to be overmassive with respect to their host galaxies \citep[e.g.,][]{harikane_jwstnirspec_2023, maiolino_small_2024}, in relation to what is expected from the local $M_{\rm BH} - M_\ast$ relation \citep{magorrian_demography_1998,RV2015}, although the extent of this effect is unclear, as other physical mechanisms could contribute to the Balmer-line broadening \citep[e.g.,][]{Rusakov2025}. In addition, LRDs can show complex morphologies in the rest-frame UV, in spite of their compactness \citep[e.g.,][]{Rinaldi2025_notdot}.

By selection most LRDs lie at $z>4$ and are typically hosted by low-mass galaxies, which sometimes are not even detected \citep[e.g.,][]{Chen2025}. Nonetheless, their metallicities do not stand out for being lower than average at high $z$,  with the exception of a few known cases with $Z \sim 0.1 \, \rm Z_\odot$ \citep{Taylor2025, Tripodi2025}, including QSO1 behind the lensing cluster Abell 2744 \citep{furtak_jwst_2023,deugenio_qso1_2025, Maiolino2025}. This is also the case for other kinds of AGN at high redshifts, whose hosts are rarely metal poor  \citep[e.g.,][]{Matsuoka2018,Mignoli2019,ubler_ga-nifs_2023}.

In LRD studies, galaxies with prominent emission lines driving the V-shaped SED colours are typically excluded \citep[e.g.,][]{Kocevski2025}. This is because the main goal is to select sources with a steeply rising rest-frame optical continuum which, in combination with the compactness criterion, allows for the identification of AGN candidates. We will argue that those discarded objects, which hereafter we call {\it pseudo-LRDs},  may merit attention on their own, as some of them could also contain AGN that are normally not accounted for in AGN demographic studies.  

In this paper we present the analysis of a highly magnified low-mass galaxy, found behind the lensing  cluster A370 at $z=5.96$, which we name {\it Pseudo-LRD-NOM} (for {\it pseudo-LRD with no metal lines}) hereafter. This source stands out for having a rather unusual spectrum: it is characterised by the presence of a very prominent $\rm H\alpha$ emission line, a much fainter but still significantly detected $\rm H\beta$ line, but no [OIII]$\lambda 5007$ or any other metal lines. In addition, the rest-frame UV spectral slope $\beta_{\rm UV}^{\rm spec}$ is quite high (low in modulus), which is indicative of significant dust attenuation.

The paper is organized as follows. In \S\ref{sec:data} we describe the utilised datasets, while in \S\ref{sec:pseudolrd} we justify the motivation for the pseudo-LRD classification of our object. In \S\ref{sec:source} we present the spectroscopic and photometric properties of {\it Pseudo-LRD-NOM}, and investigate under which physical conditions all those properties can simultaneously be explained. In \S\ref{sec:disc} we discuss our findings and compare some main properties of  {\it Pseudo-LRD-NOM} and real LRDs. Finally, in \S\ref{sec:concl} we present some concluding remarks. Throughout this paper, we consider a cosmology with $H_{0} = 70\; \rm km\;s^{-1}\;Mpc^{-1}$, $\Omega_{M} = 0.3$, and $\Omega_{\Lambda} =0.7$, and a \citet{chabrier_galactic_2003} initial mass function (IMF).

\tabletypesize{\scriptsize}
\begin{deluxetable*}{cccccccc}
\tablecaption{Coordinates and main spectral line measurements for {\it Pseudo-LRD-NOM}. \label{tab:source}}
\tablewidth{0pt}
\tablehead{
\colhead{Name} & \colhead{R.A. (J2000)} & \colhead{Dec. (J2000)} & \colhead{$z_{\rm spec.}$} & \colhead{$f \rm (narrow \, H\alpha)^a$} & \colhead{$f\rm (broad \, H\alpha)^a$} & \colhead{$f\rm (H\beta)$$^{a,b}$} & \colhead{$f \rm([OIII]\lambda5007)$$^a$} \\
& & & & FWHM~$\rm (narrow \, H\alpha)^c$  & FWHM~$\rm (broad \, H\alpha)^c$ & FWHM~$\rm (H\beta)^{c, d}$ &}
\startdata
\\
Pseudo-LRD-NOM & 02:39:56.12 & -01:34:25.19 & 5.96 & $969.6^{+105.9}_{-82.3}$ & $332.4^{+124.0}_{-145.4}$ & $88.3^{+13.2}_{-11.6}$ & $<23.2$  \\
& & & &  $741^{+143}_{-126}$ & $3214^{+841}_{-1005}$ & $741^{+143}_{-126}$ &
\\
\enddata
\tablecomments{$a$. Line fluxes are given in units of $10^{-20}\rm erg \, cm^{-2} s^{-1}$ and are not corrected for magnification;  
$b$. A single narrow component is fitted to the H$\beta$ line profile. The expected broad component is below the noise level (Fig.~\ref{fig:linefit}); 
$c$. The values of the FWHM are given in units of km/s;
$d$. By default the \textsc{unite} routine adjusts the same FWHM for the narrow components of H$\alpha$ and H$\beta$.}
\end{deluxetable*}

\section{Datasets}
\label{sec:data}

\subsection{The lensing cluster A370}
\label{sec:source}

The source analysed here is gravitationally lensed by the massive galaxy cluster Abell370 (A370) at $z=0.375$.  This cluster is one out of the six targets included in the Hubble Space Telescope (HST) Frontier Fields program \citep{Lotz2017,Bradac2019}. Its background sources have been spectroscopically followed up from the ground in the pre-JWST era \citep[e.g.,][]{Richard2021}. With JWST, A370 and its background have been observed in a number of programs, including the Cycle~1 GTO CANUCS (P.I.: C. Willott, PID \#1208); Cycle~2 GO program MAGNIF (P.I.: F. Sun; PID \#2883); and program PID \#3538 (P.I.: E. Iani). 

The increasing data availability has allowed for a continuous improvement of the A370 lensing model. The latest versions, which incorporate JWST data, are those by \citet{Gledhill2024} and \citet{Diego2025}. Here we adopt the magnification factor derived in the latter, i.e. $\mu = 15.7 \pm 0.4$.

\subsection{JWST Spectroscopy}

We analysed the source JWST spectrum taken with the Near Infrared Spectrograph's Micro-Shutter Array \citep[NIRSpec/MSA;][]{ferruit_near-infrared_2022}  as part of the JWST Cycle 1 GTO program CANUCS (P.I.: C. Willott, PID \#1208). This spectrum covers the wavelength range $0.6–5.3 \, \rm \mu m$  and has been obtained with the NIRSpec low-dispersion PRISM ($\rm R = 30–300$). The spectrum was retrieved from the DAWN JWST Archive (DJA) and detailed data reduction steps can be found in \citet{Sarrouh2026}.

\subsection{Photometric Data}

We retrieved and reprocessed all the NIRCam imaging available for A370. This includes imaging from the JWST programs CANUCS (P.I.: C. Willott, PID \#1208); MAGNIF (P.I.: F. Sun; PID \#2883); JUMPS (P.I. C. Withers; PID \#5890); VENUS (P.I. S. Fujimoto; PID \#6882); and programs PID \#3538 (P.I.: E. Iani) and PID \#5324 (P.I.: J. Pierel). For the data reduction, we employed our own custom NIRCam pipeline, which has already been tested and successfully applied in other JWST studies, including \citet{rinaldi_midis_2023}, \citet{navarro-carrera_constraints_2024}, \citet{iani_midis_2024-1}, and \citet{caputi_midis_2024}. Our software is based on the JWST pipeline version 1.20.2 \citep{Bushouse2023}, with the Calibration Reference Data System (CRDS) pipeline mapping (pmap) version 1471. It also includes extra steps similar to those described by \citet{Bagley2023}, including corrections for $1/f$ noise, stray-light effects (wisps), and residual cosmic rays (snowballs).

As our source is compact, we measured  $0^{\prime\prime}.3$-diameter aperture photometry and applied aperture corrections. We validated our photometry for {\it Pseudo-LRD-NOM} against the photometry provided in the publicly available  CANUCS  photometric catalog \citep[Data Release 1; see][]{Sarrouh2026}, for the common passbands. We also considered the available HST photometry in this catalogue, corresponding to images that have been obtained from the HST CLASH and Frontier Fields programs \citep{Postman2012, Lotz2017}. We note that the choice of aperture, between 0$^{\prime\prime}$.3 and 0$^{\prime\prime}$.7-diameter (corrected to total), and Kron apertures, does not affect the outcome of our spectral energy distribution (SED) fitting analysis (\S\ref{sec:sed}).

\begin{figure}[h!]
    \centering
    \includegraphics[width = 0.40 \textwidth]{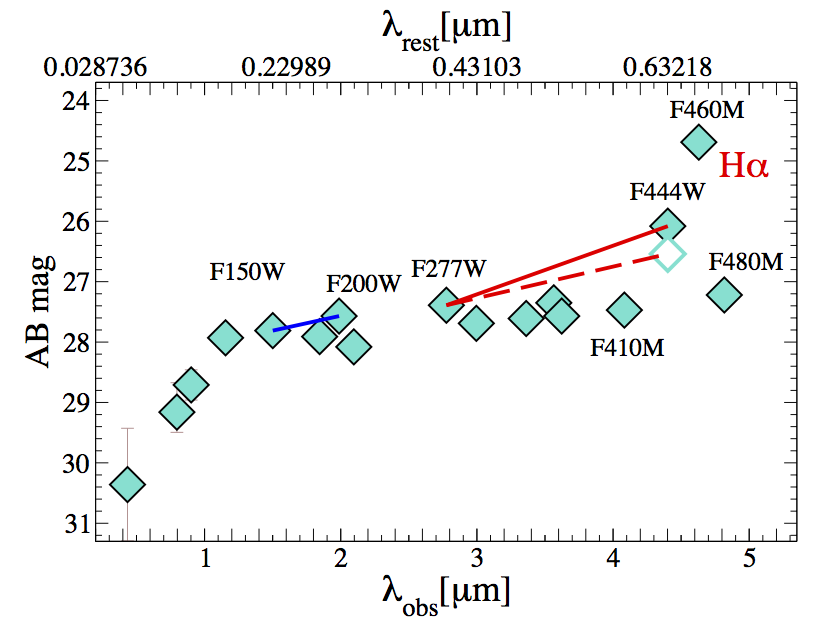}
    \caption{Observed photometry of {\it Pseudo-LRD-NOM}, uncorrected for magnification. For the F444W passband we show two data points, one corresponding to the original photometry (filled diamond) and another one obtained after correcting for the (minimum) flux density excess produced by the prominent H$\alpha$ emission line over the continuum (empty diamond). The lines join the datapoints at the passbands that are typically used to define the LRD colour selection at $z\sim6$: F115W-F200W  (blue) and F277W-F444W (red solid and dashed for the original and H$\alpha$-corrected F444W photometry, respectively). Even applying the minimum photometric correction for the H$\alpha$ line, the source moves out of the LRD selection criterion defined in \citet{Kocevski2025}.} 
    \label{fig:obssed}
\end{figure}

\section{The source {\it Pseudo-LRD-NOM}}
\label{sec:pseudolrd}

\begin{figure*}[ht!]
    \centering
    \includegraphics[width = 0.98 \textwidth]{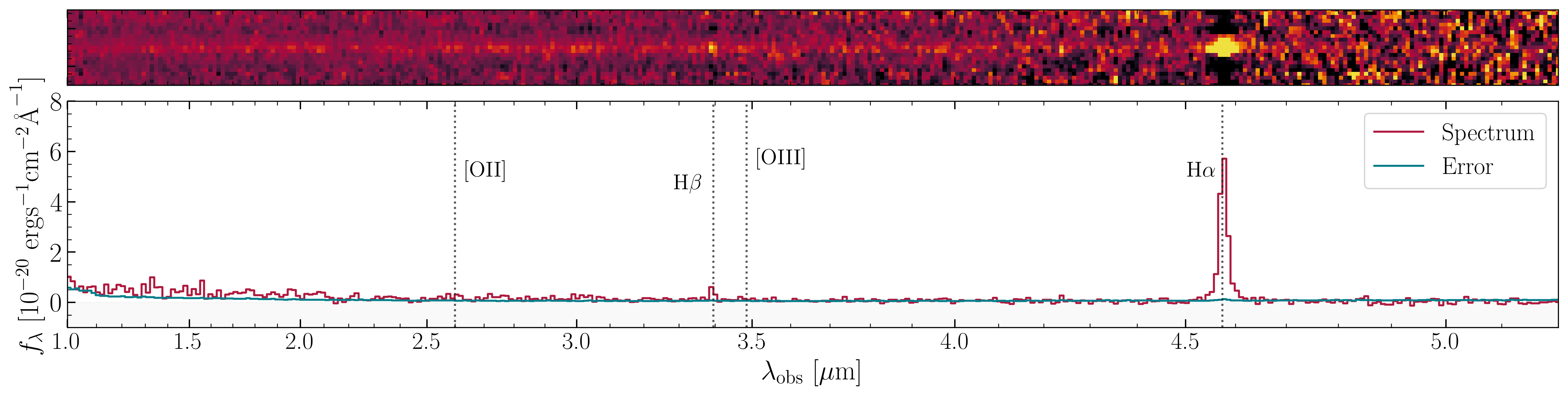}
    \caption{NIRSpec PRISM spectrum of {\it Pseudo-LRD-NOM}. While H$\alpha$ and H$\beta$ are clearly visible, no other emission lines are detected.} 
    \label{fig:spec}
\end{figure*}

Our target source appears red and compact in the JWST NIRCam images (Fig.~\ref{fig:stamps}). Its red colour is boosted by the presence of a prominent H$\alpha$ line, which we analyse in \S\ref{sec:source}. If we subtract from the F444W photometry the flux density excess produced by H$\alpha$ over the continuum, we see that the source has a more moderate red colour, as we show in Fig.~\ref{fig:obssed}. We computed the corrected F444W photometry by adding 0.46~mag  to the originally measured magnitude. This minimum correction corresponds to the lower limit of the  total H$\alpha$  rest-frame  $\rm EW_0 \gsim 800 \,\AA$ derived from the NIRSpec PRISM spectrum (see \S\ref{sec:source} and Fig.~\ref{fig:spec}), as obtained using the formula provided by \citet{MarmolQ2016}. 

The measured colours are: F150W-F200W =$-0.27 \pm 0.09$  and F277W-F444W =$1.31 \pm 0.07$ (before correction for H$\alpha$), changing to F277W-F444W~$\lsim 0.85$ (after correction for H$\alpha$). So, formally, the source would be classified as an LRD if no H$\alpha$ flux-density excess correction were applied,  as it has SED slopes  $\beta_{\rm UV}^{150-200}=-2.74\pm0.09$ and  $\beta_{\rm opt.}^{277-444}=0.62 \pm 0.07$. Once the minimum correction for H$\alpha$ is introduced, we get $\beta_{\rm opt.}^{277-444}<-0.30$.   This means that the source does not satisfy the LRD-selection criterion proposed by \citet{Kocevski2025}, while it would still satisfy the less strict criteria adopted by other authors \citep{kokorev_census_2024,Rinaldi26_jades}. Our source would rather be classified as a {\em little blue dot (LBD)} following the criteria proposed by \citet{Brazzini2026}.

Given the fact that the prominent H$\alpha$ emission line is the main driver of the red (F277W-F444W) colour of our source, we adopt the term `pseudo-LRD' for it. Nonetheless, we highlight that, independently of the formal definition, the important point is that our source is very compact and harbours a BLAGN, as LRDs and (at least some) LBDs do \citep{MM2026}. 

In addition, the large photometric differences observed between the F410M and F460M filters, and then F460M and F480M, suggest that $\rm EW_0(H\alpha)$ is significantly larger than what is inferred from the spectral lower limit previously discussed. Applying the same formula quoted above from \citet{MarmolQ2016}, we derive that the total H$\alpha$ line has $\rm EW_0(H\alpha) \approx (1200 \pm 300) \, \AA$.  This would imply that {\it Pseudo-LRD-NOM} is dominated by very young stellar populations \citep{Prieto2025}.

We present a complete SED analysis of {\it Pseudo-LRD-NOM} in \S\ref{sec:sed} and provide its coordinates in Table~\ref{tab:source}.

\section{Spectral and Photometric Analysis of {\it Pseudo-LRD-NOM}}
\label{sec:source}

\subsection{Balmer line profiles}
\label{sec:lineprof}

Figure~\ref{fig:spec} shows the NIRSpec PRISM spectrum of our target. The Balmer H$\alpha$ and H$\beta$ emission lines are clearly visible, while no other lines are detected.   The integrated H$\alpha$ line has $\rm S/N > 25$, while $\rm S/N (H\beta)\approx 8$. Given this, the lack of [OIII]$\lambda5007$ emission is immediately striking.

\begin{figure}[h]
    \centering
    \includegraphics[width = 0.45 \textwidth]{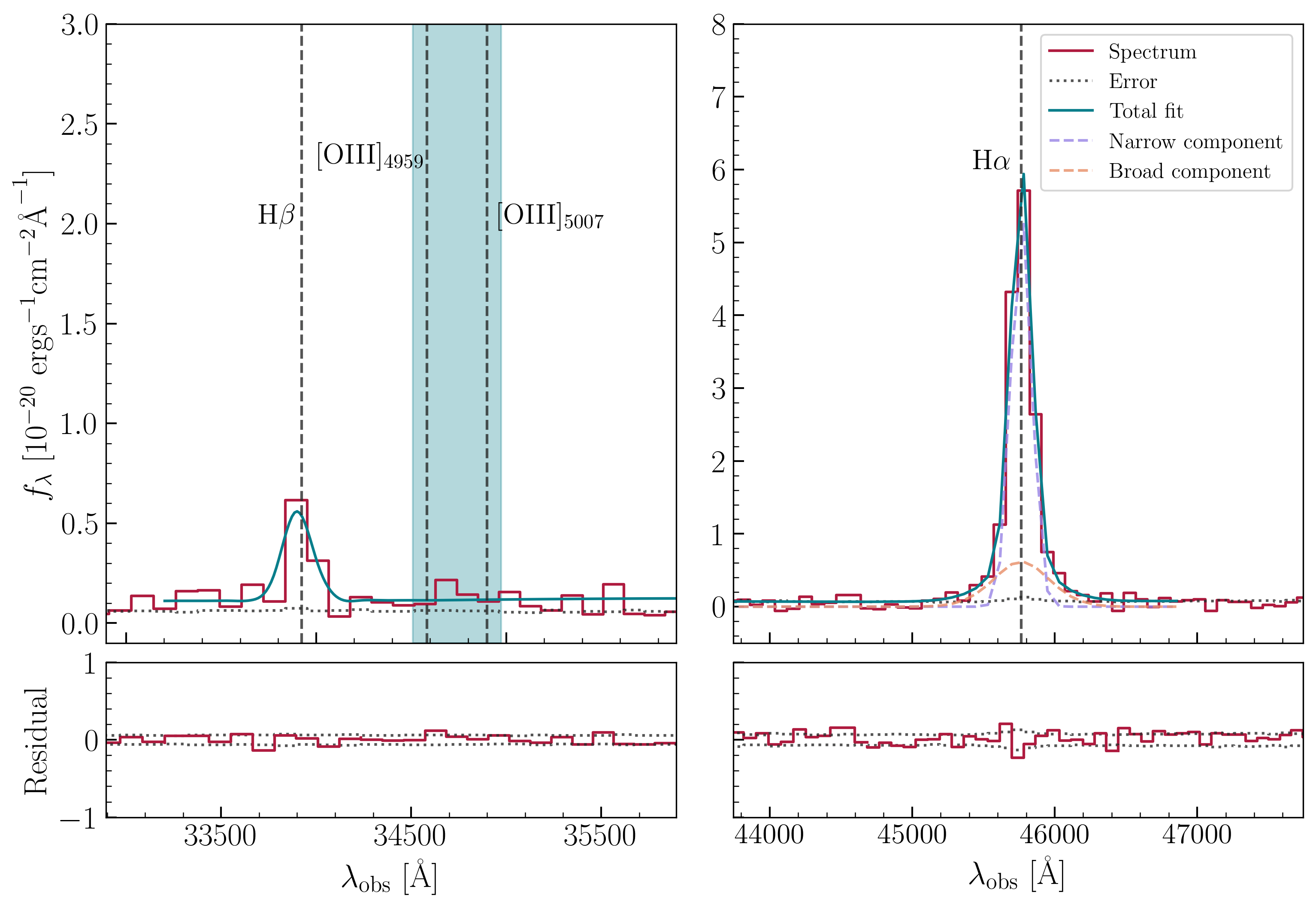}
    \caption{Zoomed-in views of the H$\beta$ and H$\alpha$ spectral lines (left and right panels, respectively). While a narrow and a broad component are clearly detected for H$\alpha$, only a narrow component is fitted for H$\beta$, with the expected broad component lying below the noise level.} 
    \label{fig:linefit}
\end{figure}

We fitted the profiles of the H$\alpha$ and H$\beta$ lines using the Python package \textsc{unite} v3.0.0 \citep{Hviding2025}. To correct for the line spread function (LSF), we used the NIRSpec built-in LSF data \citep{degraaff2024}, which takes into account the position of the object in the NIRSpec slitlet, as well as the source morphology. Our results indicate that the H$\alpha$ line has two components, a main narrow one and another fainter and broader (Fig.~\ref{fig:linefit}).  The line profile fitting with a Gaussian broad component is preferred, but considering either a Lorentzian or exponential profile for the broad component yields very similar results. This is based on the Watanabe–Akaike information criterion \citep[WAIC;][]{Watanabi2015}, for which we get $\Delta \rm (WAIC)<1$ between the different two-component models. Instead, based on the same criterion, the single component models can be discarded, as they yield $\Delta \rm (WAIC)\gsim 16$ with respect to the two-component ones.

The H$\beta$ line is best fitted solely with a narrow component, as the expected broad component lies below the noise level.  All our measured line fluxes are given in Table~\ref{tab:source}.

For the broad component of the H$\alpha$ line we measured a full width at half maximum (FWHM) of $3214^{+841}_{-1005} \, \rm km \, s^{-1}$. This large broadening indicates the presence of an active black hole at the centre of our source, as is the case for basically all LRDs \citep[e.g.,][]{Hviding2025}.  Note that a symmetric broadening of an emission line can be due to other factors, particularly clump rotation \citep[e.g.,][]{Cooper2025}. However, the large value of the H$\alpha$ broad-component FWHM makes the clump-rotation hypothesis unlikely and, crucially, no clumps are visible in the high spatial resolution NIRCam images of {\it Pseudo-LRD-NOM}. We can also discard that the line broadening is produced by other effects related to stellar kinematics, because such broadenings are typically much smaller than the value measured here for {\it Pseudo-LRD-NOM'}s H$\alpha$ emission line \citep[e.g.,][]{Tenorio2010,Amorin2024}. 

Based on the  H$\alpha$-line broad component, we can obtain an estimate of the central black-hole mass, following the prescription by \citet{GreeneHo2005}. Considering the line FWHM and magnification-corrected flux, we get $M_{\rm BH} = 5.6^{+2.9}_{-3.4} \times 10^6 \,\rm M_\odot$. The error budget in this mass estimate is dominated by the error in the H$\alpha$ broad-component FWHM, given the functional dependence on it in the \citet{GreeneHo2005} formula. This means that {\it Pseudo-LRD-NOM} hosts a massive black hole, but amongst the smallest typically found in LRDs at these redshifts, which can reach up to $\rm \sim 10^9 \, \rm M_\odot$ \citep[e.g.,][]{Maiolino2025_Chandra, Taylor2025}.

In turn, the flux measured on the narrow component of H$\alpha$ can be used to obtain an estimate of the ongoing star formation rate (SFR) in the galaxy, under the assumption that existing calibrations to obtain SFRs from $L\rm(H\alpha)$ for high-$z$ galaxies are applicable to {\it Pseudo-LRD-NOM}. As we will show in \S\ref{sec:sed}, the minimum colour excess on the stellar models that can satisfactorily explain the spectral continuum of this source is $\rm E(B-V) \approx 0.18$. Assuming the dust attenuation law by \citet{calzetti_dust_2000}, which is adequate for high-$z$ star-forming galaxies \citep{McLure2018}, we find that the magnification- and dust-corrected $L(\rm narrow \, H\alpha) \gsim (4.2^{+0.46}_{-0.36}) \times 10^{41} \, \rm erg \, s^{-1}$. To convert this minimum H$\alpha$ luminosity into an SFR, we considered the formula provided by \citet{Shapley2023} which is valid for low metallicities. We obtain SFR(H$\alpha$)$\gsim (0.90^{^+0.10}_{-0.08}) \, \rm M_\odot yr^{-1}$, which is very similar to the value obtained if we consider the conversion formula provided by  \citet{Theios2019} instead: SFR(H$\alpha$)$\gsim (0.96^{^+0.11}_{-0.09}) \, \rm M_\odot yr^{-1}$. 

\subsection{Line ratios and their implications}

\subsubsection{Interpretation of the line ratios}

The measured fluxes for the narrow component of H$\alpha$ and the total H$\beta$ imply a Balmer decrement $\rm H\alpha / H\beta =11.0^{+2.0}_{-1.7}$, which is substantially higher than the value expected for a typical warm gas cloud with no dust, following case B recombination \citep{osterbrock_astrophysics_2006}. High values of the  $\rm H\alpha / H\beta$ ratio are commonly found amongst LRDs \citep[e.g., ][]{degraaff2025,deugenio_jades_2025,furtak2025,nikopoulos2025}. A priori, such a high Balmer decrement is expected to be associated with a high dust attenuation of the warm gas. If we assume, for example, the dust reddening law by \citet{calzetti_dust_2000}, we obtain that a colour excess $\rm E(B-V) \approx 0.88$ would be necessary to explain the observed $\rm H\alpha / H\beta$ ratio solely via dust attenuation.  A very similar value would be obtained if we considered instead a `gray' attenuation curve, as that proposed in, e.g., \citet{Markov2025}, as the differences with the \citet{calzetti_dust_2000} reddening law are only significant in the UV. In any case, the choice of reddening law is incidental: none of our main results or conclusions significantly depend on the considered prescription.

However, simply invoking dust attenuation is not sufficient to explain why no metal lines are detected in the spectrum of {\it Pseudo-LRD-NOM}. While dust attenuation could explain most of these non-detections, the lack of [OIII]$\lambda5007$ in particular is the key to infer that the situation is more complex.  If the presence of dust attenuation were the only cause of [OIII] being undetected in the spectrum, then under standard conditions H$\beta$ should be undetected as well. In the following we analyse plausible solutions to this problem.

Before any further analysis, though, there is a possible caveat to be considered. It could be that the flux of the H$\beta$ line as it is observed comes partly or totally from the broad-line region, but appears as having only a narrow component because all flux beyond a central narrow window around the line center lies below the noise level. These considerations could change the narrow $\rm H\alpha / H\beta$ value quoted above. However, while at least a part of the H$\beta$ flux comes from the host galaxy (or a narrow-line region), the problem remains the same: the lack of [OIII]$\lambda5007$  emission in the galaxy spectrum  is a priori difficult to be reconciled with a simple dust-attenuation explanation.

\begin{figure}[h]
    \centering
    \includegraphics[width = 0.45 \textwidth]{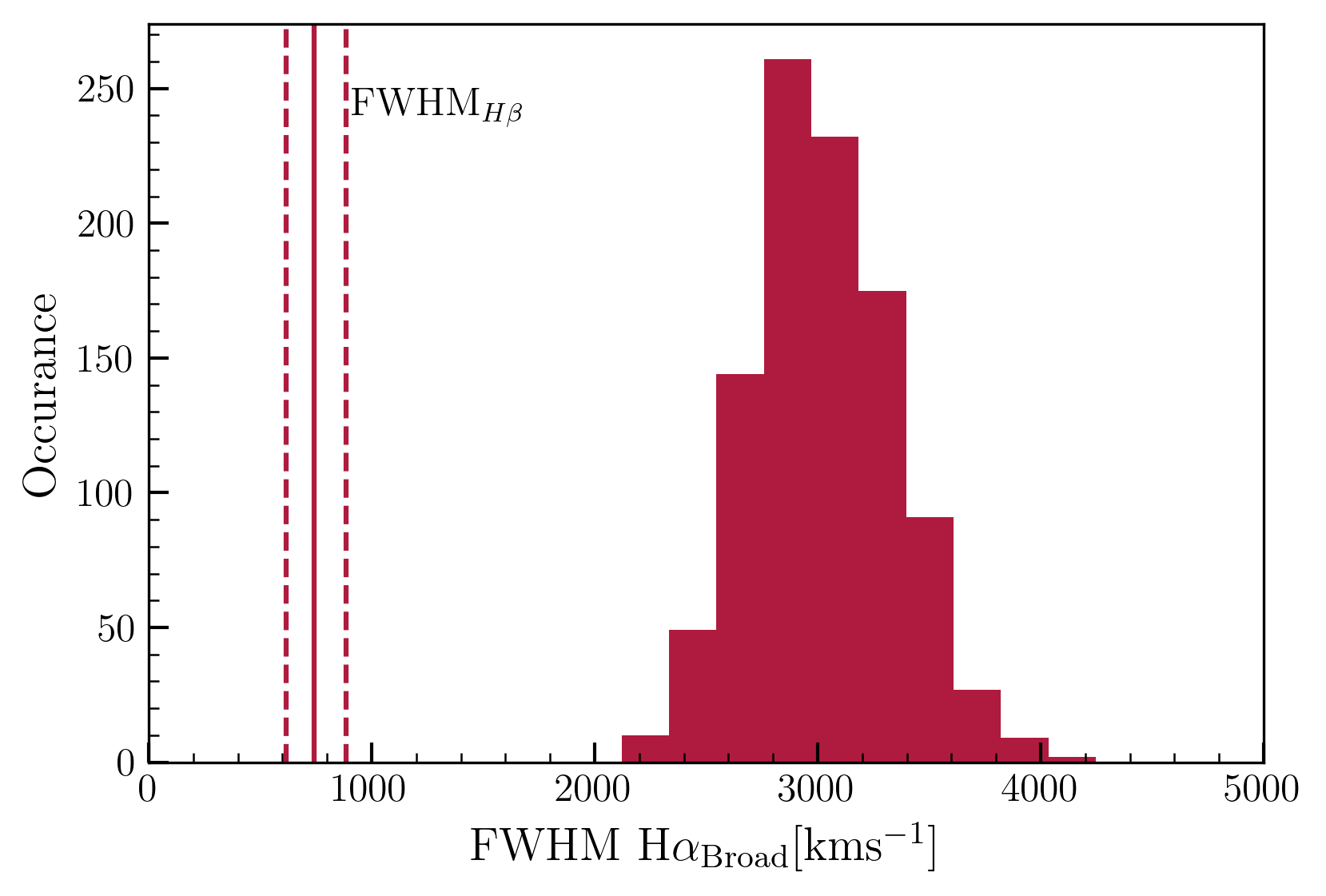}
    \caption{FWHM distribution for 1000 mock realizations of the broad H$\alpha$ line obtained after renormalizing its peak to the peak flux of H$\beta$ and degrading its S/N ratio to that observed in the H$\beta$ spectral window. In all cases these FWHM are much larger than that observed for H$\beta$. This means that H$\beta$ is genuinely dominated by a narrow component.} 
    \label{fig:mc}
\end{figure}

Instead, if all the observed H$\beta$ flux were part of the broad-line component and there is no flux associated with the narrow component of H$\beta$, then the lack of [OIII]$\lambda5007$ detection would not be puzzling, but rather a simple consequence of dust attenuation. To investigate if this could be the case, we performed a set of Monte Carlo simulations, as follows: we considered the broad component of H$\alpha$ and renormalized it such that its peak flux density value coincided with the H$\beta$ flux density peak. We also rebinned the profile to account for the different spectral resolutions of the NIRSpec PRISM at the H$\alpha$ and H$\beta$ observed wavelengths. Then we made 1000 mock realizations of the H$\alpha$ broad line component by assigning to each spectral element a random flux density value obtained from a Gaussian distribution centred at the real (renormalized) flux density, with an r.m.s. given by the noise level in the H$\beta$ spectral window. For each resulting mock broad line we measured the FWHM. Our results are presented in Fig.~\ref{fig:mc}. The obtained distribution of FWHM for the simulated broad line components has a median/r.m.s. $\rm FWHM(H\alpha)= (3108 \pm 156) \, \rm km \, s^{-1}$, which is consistent within the errors with the FWHM of the real H$\alpha$ broad component ($3214^{+841}_{-1005} \, \rm km \, s^{-1}$). The measured FWHM(H$\beta$) is $>10 \sigma$  away. Therefore, we conclude that the observed H$\beta$ line is genuinely dominated by the narrow component. In all the following we will then consider that $f(\rm H\beta)$ = $f (\rm narrow \, H\beta)$.

\subsubsection{The H$\alpha$/H$\beta$ vs. [OIII]$\lambda$5007/H$\beta$ diagram}

\begin{figure}[h]
    \centering
    \includegraphics[width = 0.50 \textwidth]{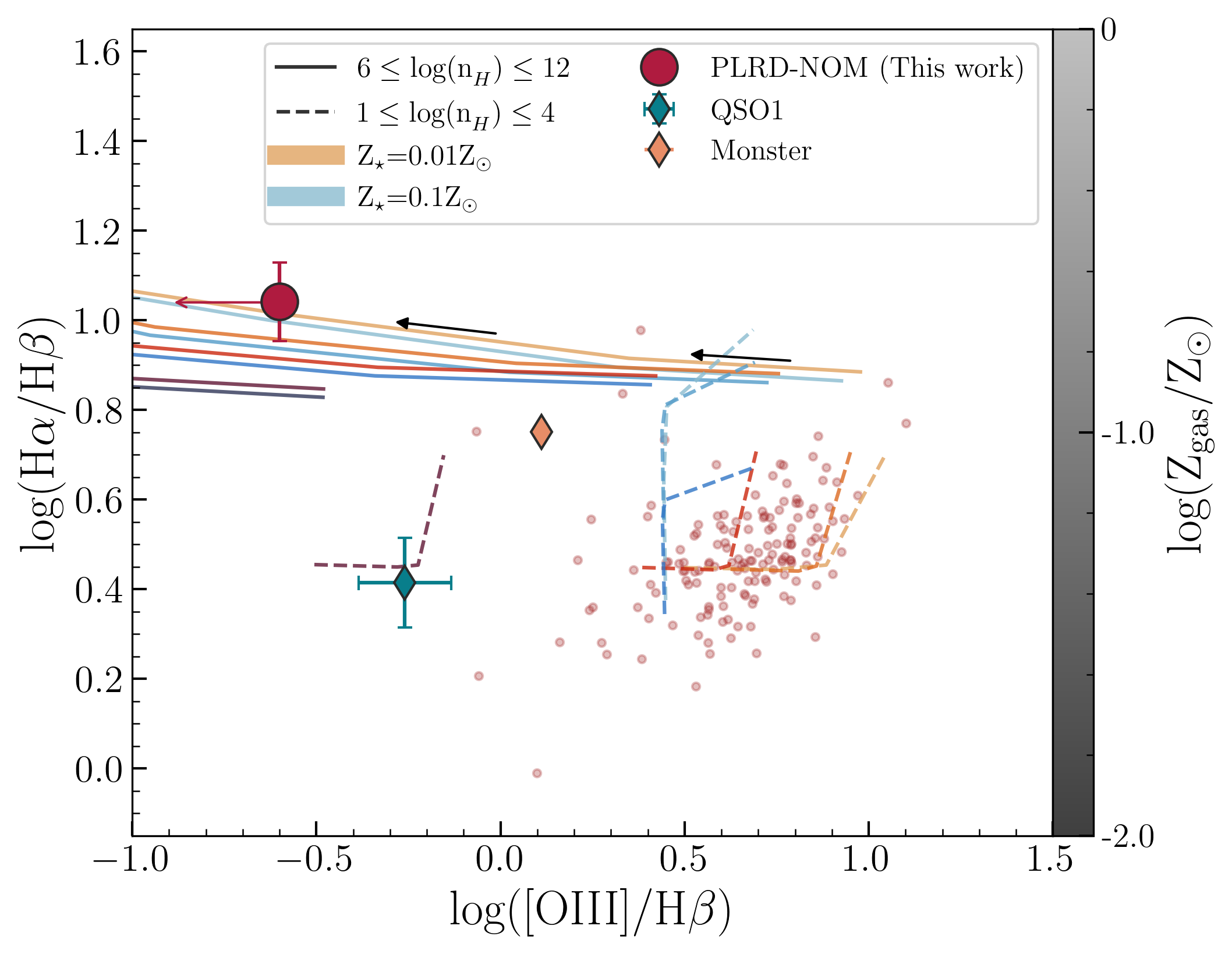}
    \caption{Balmer decrement H$\alpha$/H$\beta$ versus [OIII]$\lambda 5007$/H$\beta$ for our source {\it Pseudo-LRD-NOM} (red filled circle) and LRDs from the literature: the Monster  \citep[orange diamond; ][]{greene_uncover_2024, Labbe2025_uncover} and QSO1 \citep[teal diamond; ][]{furtak2025,deugenio_qso1_2025, Maiolino2025}, both behind Abell 2744. The small orange circles represent a general sample of low stellar-mass ($M_\ast<10^9 \, \rm M_\odot$) galaxies at $z>4$ (Cooper et al., in prep.). The solid lines show the line ratios predicted by \textsc{Cloudy} models for a starburst with age $1 \, \rm Myr$, and a range of gas and stellar metallicities, as indicated in the intensity bar and inset label.  The tracks move along the gas volume density values $n_{\rm H}$, in the range specified in the inset label.} 
    \label{fig:lineratios}
\end{figure}

In Fig.~\ref{fig:lineratios} we show the (narrow) H$\alpha$/H$\beta$ versus [OIII]$\lambda5007$/H$\beta$ ratios for our source. For context we also plot the ratios corresponding to a general sample of low-mass galaxies at $z>4$ (Cooper et al., in prep.) and a few LRDs from the literature \citep{furtak_jwst_2023,greene_uncover_2024,deugenio_qso1_2025,Maiolino2025}. From this figure we can see that the vast majority of low-mass galaxies follow a broad correlation in this diagram, as it is expected from the canonical situation in which the oxygen abundance (which is one of the main parameters governing the [OIII]/H$\beta$ ratio) increases with dust attenuation, as given by the H$\alpha$/H$\beta$ ratio. Instead, some LRDs appear to be off this relation. This is particularly evident for our source {\it Pseudo-LRD-NOM}, which strongly suggests that the physical conditions are different in these objects. 

The upper limit on the [OIII]/H$\beta$ ratio ($<0.25$) can be used to obtain an estimate of the upper limit of the gas metallicity  12+$\rm \log_{10}(O/H)$. Although in principle the relation between these two quantities is bivariate, the upper metallicity branch can safely be discarded by the lack of metal lines in the spectrum. Considering the recent calibration obtained by \citet{Cataldi2025}, for {\it Pseudo-LRD-NOM} we obtain 12+$\rm \log_{10}(O/H) < 7.0$, i.e. $<0.02 \, \rm Z_\odot$ (this value would be even lower considering other calibrations from the literature). So this would suggest that the ISM warm gas in {\it Pseudo-LRD-NOM}'s host is close to pristine condition.  However, these calibrations are based on galaxies whose ISM's densities are less extreme than for {\it Pseudo-LRD-NOM}, so their application to LRDs may not be straightforward, as we discuss below.

We produced a range of line emission models using the software \textsc{Cloudy} \citep{ferland_2013, ferland_2017} to investigate under which conditions the observed line ratios of {\it Pseudo-LRD-NOM} can be reproduced. For our ionizing source, we consider stellar population models from BPASS \citep{eldridge_binary_2017} that include a burst in star formation and binary systems. We consider both solar and sub-solar metallicities for the stellar and gas-phase metallicities and employ stellar populations with ages in the range $10^6$ to $10^8$~years. These stars are the only source of ionizing photons and no dust is present in any of our models.

 Although we will argue in \S\ref{sec:sed} that {\it Pseudo-LRD-NOM} has actually non-negligible dust attenuation of its spectral continuum, the main objective here is to show that the spectral line ratios can be explained even without invoking the presence of dust. We will discuss later the implications of the high Balmer decrements being partly produced by dust attenuation (see \S\ref{sec:sed}).

In Fig.~\ref{fig:lineratios} we show the \textsc{Cloudy} model tracks corresponding to a very young starburst with age 1~Myr.  We see that only models with high gas density ($n_{\rm H}> 10^6 \, \rm cm^{-3}$) can reproduce {\it Pseudo-LRD-NOM}'s line ratios. Different combinations of stellar and gas metallicities between $0.01 \, \rm Z_\odot$ and $0.1 \, \rm Z_\odot$ are similarly suitable.  Note that the high gas densities needed for our source are above the [OIII]$\lambda 5007$ critical density, which is $7 \times 10^5 \, \rm cm^{-3}$ \citep{osterbrock_astrophysics_2006, BaskinLaor2005}. This means that, in {\it Pseudo-LRD-NOM}, the [OIII]$\lambda 5007$ line emission would mainly be governed by collisional excitation rather than photoionization.

\begin{figure*}[ht!]
    \centering
    \includegraphics[width = 0.95 \textwidth]{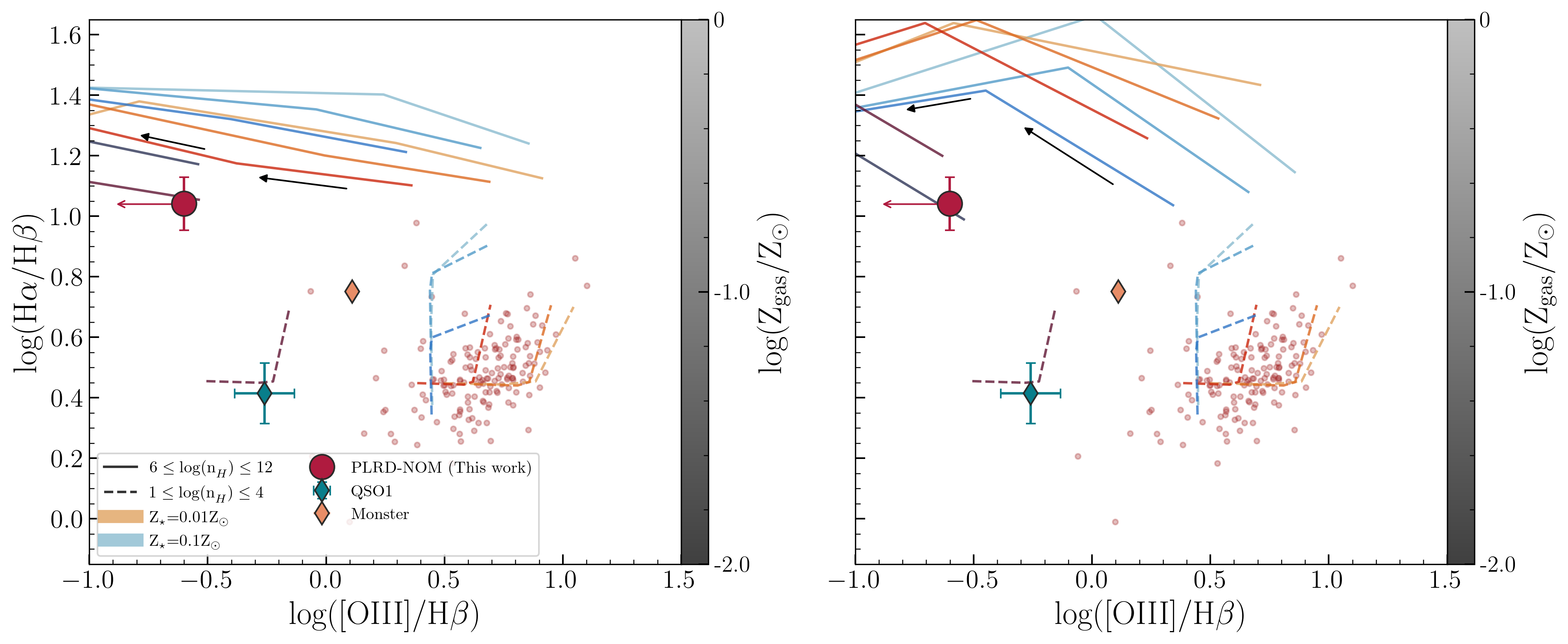}
    \caption{The same as Fig.~\ref{fig:lineratios}, but showing \textsc{Cloudy} model tracks for older starbursts with ages 10~Myr ({\it left}) and 100~Myr ({\it right}). These plots indicate that any stellar age $\geq 10 \, \rm Myr$ simultaneously requires very low stellar and gas metallicities  to reproduce the line ratios of PLRD-NOM.} 
    \label{fig:lrolderage}
\end{figure*}

Therefore, the extremely low metallicities inferred from calibrations based on more normal galaxies should be taken with caution in this case, as very high gas densities can also be a driver of low [OIII]/H$\beta$ ratios. Deeper spectroscopic data allowing for a more stringent constraint on the [OIII]$\lambda 5007$ emission is necessary to conclude on extremely low metallicities for {\it Pseudo-LRD-NOM}. In any case we note that the \textsc{CLOUDY} model predictions do take into account the degeneracies between gas density and stellar/gas metallicities and still low metallicity values $\lsim 0.1 \, \rm Z_\odot$ are derived from them.
 
The physical conditions derived for {\it Pseudo-LRD-NOM} are in contrast with those for the vast majority of low-mass galaxies, whose line ratios can be reproduced by models of starbursts with much lower gas densities (Fig.~\ref{fig:lineratios}). At the same time, the line ratios of QSO1 can also be reproduced with models of starbursts with moderate densities, but only in combination with very low gas/stellar metallicities.

Crucially, the \textsc{Cloudy} models considered here do not incorporate any dust attenuation, which indicates that the physical conditions of the host in which {\it Pseudo-LRD-NOM} is embedded can be explained without invoking any dust. The very high densities and low metallicities which are necessary to reproduce the spectral lines of {\it Pseudo-LRD-NOM} are similar to those of nuclear star clusters or proto-globular clusters  \citep[e.g.,][]{Vanzella2023, Adamo2024}. We note that recently \citet{Yan2025} also analysed the physical conditions to explain the high Balmer decrements observed in some LRDs without the need of dust. However, these models were based on the assumption that the main radiation source is the central AGN, while here we only consider stellar emission to model {\it Pseudo-LRD-NOM}.

Fig.~\ref{fig:lrolderage} is equivalent to Fig.~\ref{fig:lineratios}, but shows \textsc{Cloudy} model tracks corresponding to older starbursts with ages 10 and 100~Myr. We see that in these two cases we can also reproduce the spectral line ratios  of {\it Pseudo-LRD-NOM} with the models of dust-free,  very dense starbursts. However, interestingly, as the age increases, only very low gas-metallicity models are suitable.

\subsection{Further clues from the spectral slope $\beta_{\rm UV}^{\rm spec}$ and SED fitting on the photometry}
\label{sec:sed}

To further understand {\it Pseudo-LRD-NOM}'s nature, we study its SED as traced by the HST and NIRCam broad-band photometry.  We considered total photometry in all passbands, but we explicitly excluded both the F444W and F460M filter flux densities, as they are significantly affected by the H$\alpha$ emission.

To perform the SED fitting,  we made use of the  code  \textsc{Cigale} \citep{boquien_cigale_2019}, considering different combinations of stellar$+$nebular and stellar$+$AGN models. For the stellar templates we employed a delayed star-formation history with a secondary burst, allowing the age of the stellar population to select from 1~Myr to 1000~Myr. Stellar populations are produced from BC03 models \citep{bruzual_stellar_2003}, with stellar metallicites varying between $\rm Z_{\odot}$ and $\sim$1/1000~$\rm Z_{\odot}$. We included the effect of dust attenuation following a modified \citet{calzetti_dust_2000} attenuation law with E(B-V) values ranging from 0 to 1. For the stellar$+$nebular model, nebular emission is included with a fixed ionization parameter $\log_{10}(\mathrm{U})=-2$ and gas-phase metallicities varying across the same range as the stellar metallicity. In the stellar$+$AGN model, we utilise the SKIRTOR 2016 AGN model \citep{stalevski_dust_2016} with an opening angle of 70$^{\circ}$ and AGN fractions of 0, 0.01, and $0.1-0.9$ with steps of 0.1.

\begin{figure*}[ht!]
    \centering
    \includegraphics[width = 0.95 \textwidth]{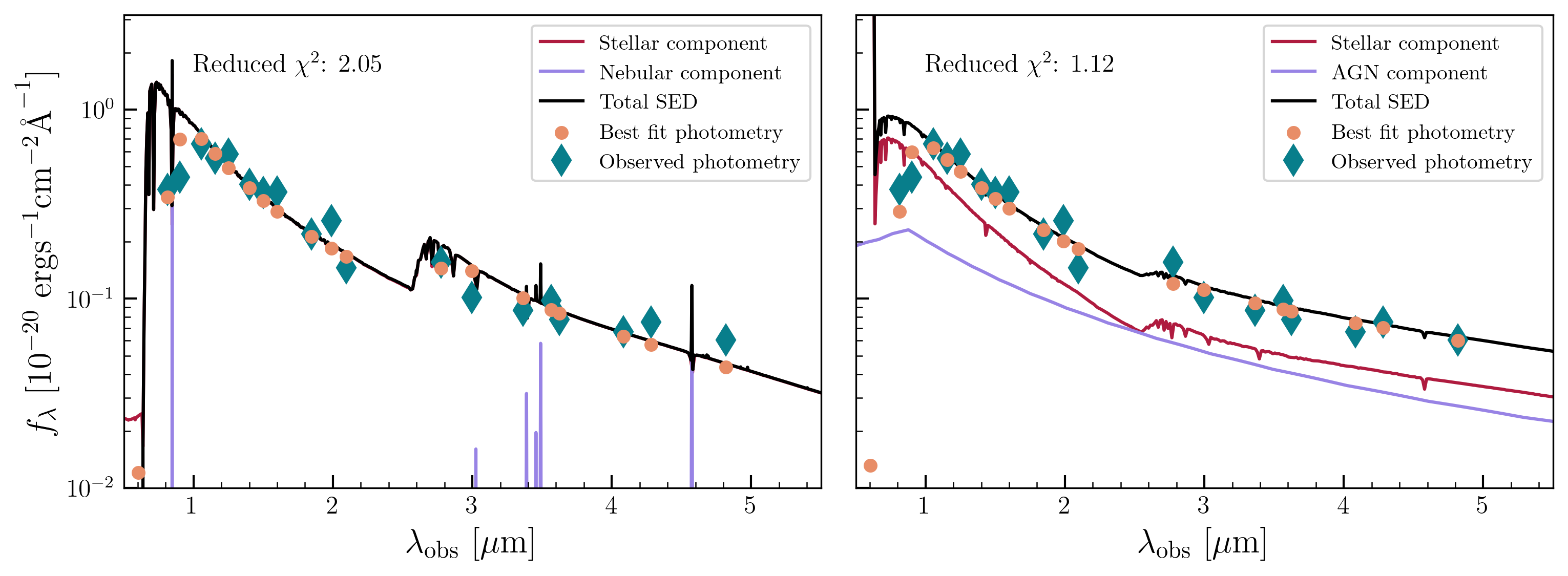}
    \caption{The best-fit SEDs obtained with \textsc{Cigale} on the source NIRCam photometry, with different model combinations. {\it Left:} stellar template + nebular emission. The nebular emission contribution is virtually insignificant. {\it Right:} Stellar template + AGN.  Although the AGN is subdominant at rest-frame UV wavelengths, it makes for about $40\%$ of the total light in the rest-frame optical. The best fit parameters are given in Table~\ref{tab:sed}. } 
    \label{fig:sed}
\end{figure*}

Fig.~\ref{fig:sed} shows the best-fit SEDs obtained using a combination of stellar and nebular templates, and a combination of stellar and AGN templates. We see that both combinations produce a  good fit to the total SED, with the combination of stellar and AGN templates slightly preferred. The derived AGN contribution to the SED is about $\approx 10-40\%$, depending on the wavelength range. In any case, both template combinations yield very similar stellar masses, indicating that the obtained stellar mass can be considered robust. Reassuringly, for any of the template combinations, the properties of the stellar component are quite similar: in both cases, the stellar metallicity is the lowest possible, the age is quite young ($\approx 44-124 \, \rm Myr$) and the colour excess is significant. Forcing \textsc{Cigale} to consider only higher metallicity models and/or higher ages reduces the output E(B-V) value, but the resulting reduced $\chi^2$ is significantly higher, indicating that these alternative solutions are not preferred.

We investigated whether this need for dust attenuation for the stellar continuum could be related to the special conditions revealed by the spectral properties of {\it Pseudo-LRD-NOM}. To do this we compared the rest-frame UV continuum slope directly derived from the spectrum (as the continuum is detected at those rest-frame wavelengths) with those produced by \textsc{Cloudy} for the models that best reproduce the spectral line ratios (Fig.~\ref{fig:lineratios} and \ref{fig:lrolderage}).  From the observed spectrum we get $\beta_{\rm UV}^{\rm spec} = -1.20 \pm 0.28$, where the error corresponds to the standard deviation obtained from bootstrapping over 500 iterations, with the spectrum being perturbed according to the error spectrum.  Instead, from the \textsc{Cloudy} models that fit the line ratios we get that the intrinsic $\beta_{\rm UV}$ values are very small ($\beta_{\rm UV} < -2$). Therefore, we conclude that a non-negligible amount of dust attenuation is also necessary to reproduce the relatively red rest-frame UV slope measured on {\it Pseudo-LRD-NOM}'s spectrum.

For the case of a 1~Myr old starburst, there are multiple \textsc{Cloudy} tracks that reproduce our {\it Pseudo-LRD-NOM}'s spectral line ratios. They all have $\beta_{\rm UV}$ values between -2.8 and -2.9, which implies that a colour excess $\rm E(B-V)\approx 0.4$ would be necessary to reproduce $\beta_{\rm UV}^{\rm spec} = -1.20$ measured on {\it Pseudo-LRD-NOM}'s spectrum (assuming the \citet{calzetti_dust_2000} dust-attenuation law; see e.g., \citet{reddy_mosdef_2018}).  For a 10~Myr old starburst, only models with very high gas density ($n_{\rm H} > 10^{10} \, \rm cm^{-3}$) and very low metallicities ($Z \approx 0.01 \, \rm Z_\odot$) can reproduce the observed line ratios. Such models have $\beta_{\rm UV} = -3.2$, so a value $\rm E(B-V)\approx0.5$ is required to flatten the rest-frame UV slope to the observed value. Finally, for the 100~Myr old starburst, the track that reproduces the line ratios has $\beta_{\rm UV} = -2.05$, which implies the need of $\rm E(B-V)\approx 0.2$ to account for the observed $\beta_{\rm UV}^{\rm spec} = -1.20$.

In the light of these findings, it is likely that the H$\alpha$/H$\beta$ line ratio is at least partly affected by dust attenuation. The colour excess values derived above imply that the effect of dust attenuation could by itself produce an H$\alpha$/H$\beta$ ratio between $\approx 3.9$ and $6.2$, but not as high as the measured H$\alpha$/H$\beta = 11.0$. Some previous works have found that, in some galaxies, the gas could have up to $\times 2$ higher attenuation in the UV than the stars \citep[e.g.,][]{Reddy2020,Shivaei2020}. If this were the case, the resulting H$\beta$ line could be up to $\approx 15\%$ more attenuated than inferred from the continuum colour excess (also assuming the \citet{calzetti_dust_2000} dust-attenuation law. Considering this, the  H$\alpha$/H$\beta$ ratio could as high as $\approx 7$. In no case, however, we can explain the observed H$\alpha$/H$\beta=11.0$ ratio directly measured from the spectrum solely based on the effect of dust attenuation.

At the same time, one has to consider that the [OIII]/H$\beta$ is virtually independent of dust attenuation. If the intrinsic (unattenuated) H$\alpha$/H$\beta$ value were very low, then {\it Pseudo-LRD-NOM} would be located close to QSO1 in Fig.~\ref{fig:lineratios}. The inferred gas densities would be lower, but also the necessary gas/stellar metallicities. And this would have to be coupled with high dust extinction to reproduce the observed line ratios for {\it Pseudo-LRD-NOM}, resulting in a more intriguing scenario, which we discuss in \S\ref{sec:disc}.

The stellar mass derived from the SED fitting is $\log_{10}{(M_\star/M_\odot)=8.55 \pm 0.20}$, which after correction for magnification becomes $\rm \log_{10}{(M_\star/M_\odot)=7.35 \pm 0.21}$  (adopting $\mu = 15.7 \pm 0.4$). Considering the dust- and magnification-corrected SFR lower-limit that we derived in \S\ref{sec:lineprof}, i.e., SFR(H$\alpha$)$\gsim 0.90^{+0.10}_{-0.08} \, \rm M_\odot yr^{-1}$, we get a specific SFR value $\rm \log_{10}\rm (sSFR/yr^{-1}) \gsim -7.40 \pm 0.44$ (which is independent of magnification). This high minimum sSFR implies that {\it Pseudo-LRD-NOM} is a starburst galaxy, following the empirical starburst definition of \citet{caputi_star_2017,caputi_alma_2021}.

Finally, the estimated half-light radius of the source, as obtained from the F150W-band image, is  $\rm (520 \pm 270) \, pc$. After correcting it for magnification (by dividing it by $\sqrt{\mu}$) it becomes $R_{1/2}= (131 \pm \, 68) \, \rm pc$.   This gives a stellar-mass surface density $\Sigma_\ast = 418^{+725}_{-310} \, \rm M_\odot / pc^2$ (independent of magnification).  This is similar to the stellar surface density found in massive star clusters \citep{Vanzella2023, Mowla2024} and even compatible within the error bars with the stellar surface density of nuclear star clusters  \citep[e.g.,][]{Neumayer2020,Fahrion2021}.

\begin{deluxetable}{ccc}
\tablecaption{Best-fit SED parameters for {\it Pseudo-LRD-NOM}. \label{tab:sed}} 
\tablehead{
\colhead{Parameter} & \colhead{Stellar + Nebular}  & \colhead{Stellar + AGN}  
}
\startdata
Reduced $\chi^2$ & 2.05 & 1.12 \\
$\log_{10}(M/\rm M_\odot) ^{a}$ &  $8.68 \pm 0.72$ & $8.55 \pm 0.20$ \\
E(B-V)$^{b}$ & $0.33 \pm 0.20$ & $0.45 \pm 0.17$ \\
Age (Myr) & $124.2 \pm 95.3$ & $44.5^{+44.5}_{-40.2}$\\
Stellar Metallicity & 0.0001 $\approx 0.7\% \rm Z_\odot^{\, c}$ & 0.0001 $\approx 0.7\% \rm Z_\odot^{\, c}$
\enddata
\tablecomments{Results at fixed $z_{\rm spec}=5.96$. a) The quoted stellar masses do not carry any magnification correction. b) The colour excess applies only to the stellar component. c) Considers that the solar metal fraction is $Z_\odot=0.014$ \citep{asplund_chemical_2009}. }
\end{deluxetable}

\section{Discussion}
\label{sec:disc}

\subsection{The complex scenario revealed by the simple spectrum of Pseudo-LRD-NOM}

The spectral properties of {\it Pseudo-LRD-NOM} are quite unusual, as they include a high Balmer decrement (narrow) H$\alpha$/H$\beta = 11.0$ simultaneously with a low ratio [OIII]$\lambda 5007$/H$\beta$$<0.25$. High Balmer decrements are relatively common amongst LRDs \citep[e.g.,][]{nikopoulos2025, Rusakov2025}. Although some first studies have  explained them simply by the presence of dust attenuation, subsequent analyses suggested  a more complicated picture  \citep[e.g.,][]{perez-gonzalez_what_2024,Setton2025}. Very recently, \citet{Yan2025} showed that high Balmer decrements can be explained via high-density gas emission from the AGN broad-line region within the LRDs, without the need of invoking dust.

The situation for {\it Pseudo-LRD-NOM} is somewhat different, as the high H$\alpha$/H$\beta$ is dominated by the narrow line components, which originate in the narrow-line region  or the galaxy host. We have not made the difference between the two in this paper, as the available spectrum does not have the necessary spatial information for this purpose. However, we note that it is unlikely that the PLRD-NOM's spectrum originates in a narrow-line region, even if the narrow components of the Balmer (H$\alpha$, H$\beta$ lines have a FWHM compatible with them. This is because typically the gas in narrow-line regions is significantly less dense than in HII regions \citep{Kakkad2018} and here we showed that the observed line ratios require the presence of very dense gas.  Moreover, the ratio between the H$\alpha$ narrow and broad-component luminosities in our source is $\approx 2.9 \pm 0.4$, which is  significantly higher than the value that would be expected if the narrow component were originated in a narrow-line region \citep[e.g.,][]{sternlaor_2012}. 

In any case, the conclusion of the gas being metal-poor would not be affected, as the relations between the main line ratios and oxygen abundances for narrow-line regions and HII regions mainly differ around and above the solar value  \citep{Zhu2024}.  In addition, although invoking dust attenuation is in principle not necessary to explain the high H$\alpha$/H$\beta$, we do infer the presence of dust from the relatively flat rest-frame UV continuum slope $\beta_{\rm UV}^{\rm spec} = -1.20$ in {\it Pseudo-LRD-NOM'}s spectrum.

\begin{figure}[h]
    \centering
    \includegraphics[width = 0.45 \textwidth]{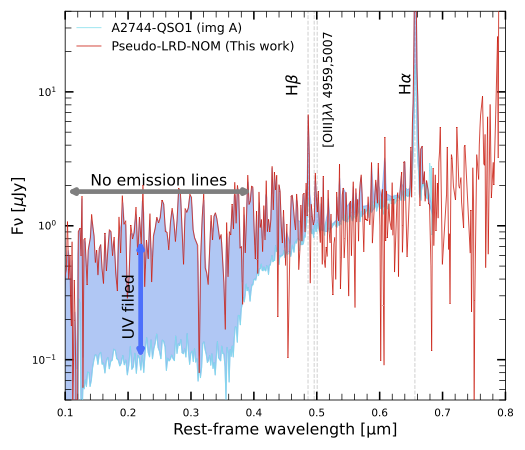}
    \caption{Comparison of the {\it Pseudo-LRD-NOM} and QSO1 spectra, renormalized at rest-frame wavelength $\approx 0.53 \, \rm \mu m$. } 
    \label{fig:bb}
\end{figure}

Fig.~\ref{fig:bb} shows a comparison of the {\it Pseudo-LRD-NOM} and QSO1 spectra. A main difference between the two is the strength of the  Balmer break, which is very evident for QSO1 \citep{Ji2025}, and absent for {\it Pseudo-LRD-NOM}. The prominent Balmer breaks observed in some LRDs do not have a stellar origin, but are rather produced by ultra-dense ($n_H \gsim \rm 10^{10} \, cm^{-3}$) and turbulent gas, most likely present in the AGN broad-line region \citep[e.g.,][]{IM2025}. This ultra-dense gas can trap the UV photons produced by the black-hole accretion, resulting in the sharp spectral drop observed at short wavelengths \citep[e.g.,][]{Liu2025}. The lack of such a sharp feature for {\it Pseudo-LRD-NOM} may be indicating that its central black hole is not surrounded by such ultra-dense gas structure. And the gas in its host galaxy, even if dense, does not reach the ultra-high densities of the gas in LRD's broad-line regions.

The lack of a significant Balmer break makes {\it Pseudo-LRD-NOM}  resemble other kinds of AGN found to be hosted by low-metallicity galaxies at high $z$ \citep[e.g.,][]{ubler_ga-nifs_2023}. In all these sources, the flux of UV photons is high enough to make the Balmer break disappear.

\subsection{An active black hole in a dense star cluster}

Interestingly, most of the light of {\it Pseudo-LRD-NOM} can be explained by having a stellar origin, with the need of only a relatively minor AGN component for the SED. The presence of an AGN becomes most evident by the identification of a broad component in the H$\alpha$ line. Although a broad-line component could in principle have a different origin, its FWHM is so large that the association with an active black hole is the most likely explanation. The presence of a BLAGN is now recognised as quintessential to the LRD nature. However,  the black-hole masses derived from the line broadening are systematically higher than what is expected from the local (and even high-$z$) $M_{BH}$ vs. $M_\ast$ relation \citep[e.g.,][]{harikane_jwstnirspec_2023,pacucci_jwst_2023}. The situation for {\it Pseudo-LRD-NOM} is similar: we find $M_{\rm BH} / M_\ast \approx 0.18$, which means that the central black hole is overmassive by 1-2 orders of magnitude. So the presence of overmassive black holes is not exclusive to LRDs, but rather also applies to other types of AGN at high $z$ \citep[e.g.,][]{Onoue2019,harikane_jwstnirspec_2023}. Alternatively, it could be that the line broadening is enhanced by physical processes other than kinematics of the broad-line region, in which case the black-hole masses could be overestimated. For example, electron scattering in dense ionized gas clouds could be partly responsible for the line broadening \citep[e.g.,][]{Laor2006,IH2025}.

{\it Pseudo-LRD-NOM'}s host appears to be a dense and dusty starburst, with a stellar density typical of massive star clusters and even nuclear star clusters. The simultaneous presence of an AGN and a nuclear star cluster in LRDs has been recently proposed via theoretical considerations \citep[e.g.,][]{Inayoshi2025}. The spectral properties of {\it Pseudo-LRD-NOM} reveal that this combination is present in this source. As a matter of fact, the connection between nuclear star clusters and black-hole growth has been investigated since at least a few decades ago \citep[e.g.,][]{PZwart2004,PZwart2006}. 

In general, it is believed that the presence of nuclear star clusters could facilitate the efficient growth of massive black holes by funneling gas to the galaxy centre \citep[e.g.,][]{Naiman2015,Partmann2025,Su2025}.  They could potentially explain the early growth of massive black holes via synchronised episodes of black-hole accretion and nuclear star formation. The identification and further study of other low-mass galaxies with nuclear activity and high stellar/gas densities are necessary to understand the details of this phenomenon.

Recently \citet{Dekel2025} proposed a mechanism for LRD formation:  star clusters merging to form compact central star clusters. Each star cluster carries a core-collapse black hole, which contributes to the formation of a central, massive black hole in the merging process.   A series of gas-rich compaction events deepen the galaxy central potential well, allowing for the retention of the central black hole. In this scenario, which is similar to that considered by \citet{Inayoshi2025}, {\it Pseudo-LRD-NOM} could represent a system which is in between compaction events. The inflow of pristine gas from the dark matter halo could keep the metallicity low until the compaction triggers new star formation activity and subsequent chemical enrichment. So, in summary, {\it Pseudo-LRD-NOM} could constitute an LRD in the making.

We note that {\it Pseudo-LRD-NOM'}s nature is in principle different to that of the local {\it blueberry galaxies} studied by \citet{Yang2017} and the sources studied by \citet{Elmegreen2017}. These galaxy populations have low masses and are metal poor, but are also significantly less compact than {\it Pseudo-LRD-NOM}. Therefore, their resulting stellar mass surface densities are also lower. In addition, by selection these galaxies are dust-free and have no reported evidence of the presence of black-hole activity.

\subsection{The simultaneous presence of extremely low metallicities and dust attenuation}

Another property inferred for {\it Pseudo-LRD-NOM}'s host is its very low metallicity: $Z = 0.01$-$0.1 \, \rm Z_\odot$ for the gas and stars from the spectral modelling,  and even lower stellar metallicities from the SED fitting. Sub-solar metallicities are the rule amongst low-mass galaxies at high $z$, and some LRD hosts have been found to have gas metallicities of $\sim 0.1 \, \rm Z_\odot$ \citep[e.g.][]{Maiolino2025, Taylor2025, Tripodi2025}.  {\it Pseudo-LRD-NOM} appears to be at an earlier stage of chemical enrichment.

It is particularly noteworthy that this source appears
to exhibit simultaneously very low metallicities and significant dust attenuation. The presence of dust in metal-poor galaxies is known at low $z$ \citep[e.g.,][]{Galliano2005, Remy2013, Izotov2014}. It has also been reported in metal-poor, star-forming galaxies at high $z$, particularly gamma-ray burst hosts \citep[][]{heintz2023_dust}.  It can be explained via a quick dust injection produced by supernovae \citep[e.g.,][]{Hirashita2002,MartGonz2022}  or carbon-enriched Wolf Rayet stars \citep{Lau2021}, before the galaxy becomes significantly metal-enriched. Early metal pollution from massive stars has recently been invoked to explain the metal abundances of higher $z$ galaxies, such as GNz11 \citep{Ebihara2026}.

Thus, {\it Pseudo-LRD-NOM} could be at a more advanced chemical enrichment stage than sources dominated by pristine gas, but still be at the initial stages of chemical enrichment. Alternatively, the low metallicity could be maintained by a continuous accretion of pristine gas from the dark matter halo in which {\it Pseudo-LRD-NOM} resides, producing metal dilution \citep[e.g.,][]{Cresci2010,Troncoso2014,Heintz2023}. The search for signatures of inflowing gas in other sources which do show metal lines in their spectra would help to test this scenario. 

The results of the SED fitting suggest that {\it Pseudo-LRD-NOM}'s host is a young starburst with age 44-124~Myr. In turn, the flux density excess (with respect to the continuum) observed in the F460M filter due to the presence of H$\alpha$ indicates that $\rm EW_0(H\alpha)\approx 1200 \, \rm \AA$. This high value also indicates that {\it Pseudo-LRD-NOM} must be dominated by young stellar populations, in agreement with the results of the SED fitting.

The presence of an early dusty phase before starbursts shine dust free has been predicted for the formation of luminous UV galaxies at early cosmic times \citep{Ferrara26}. In this scenario, the produced dust would be pushed to the galaxy outskirts once supernova-driven gas outflows become important. These gas outflows are also a natural consequence of stellar evolution. {\it Pseudo-LRD-NOM} could be to our knowledge the first reported example of such early phase of star-formation activity.

The suppression of metal lines in PLRD-NOM's spectrum is to some extent the consequence of the high-gas density and this parameter is degenerate with metallicity. This is why typical calibrations of [OIII]/H$\beta$ versus oxygen abundance would not apply for the PLRD-NOM. In any case, we note that the \textsc{CLOUDY} model predictions shown in  Fig.~\ref{fig:lineratios} and \ref{fig:lrolderage} take into account those degeneracies and from them we derive that, even at high gas densities, low stellar/gas metallicity values $\lsim 0.1 \, \rm Z_\odot$ are necessary to reproduce the spectral properties of PLRD-NOM.

A further intriguing possibility is that {\it Pseudo-LRD-NOM} could be hosting Population III (Pop III) stars. These stellar populations are in principle expected to have spectra with only emission lines produced by H and He atomic transitions \citep[e.g.,][]{nakajima_diagnostics_2022}, but some minor metal self-pollution may be possible, even while Pop~III stars still dominate the new star-forming regions \citep[e.g.,][]{Sarmento2017,Rusta2025}. To date, only a small amount of high-$z$ candidates have been found to likely host significant Pop~III star-formation activity \citep[e.g.,][]{Vanzella2023,Morishita2025, Maiolino2026}. 

The lack of detectable HeII$\lambda4686$ emission in {\it Pseudo-LRD-NOM}'s spectrum prevents us to extract such a conclusion. We note that the spectral continuum is marginally detected at the position of the HeII$\lambda4686$ emission line and, thus, the line should also be detected if such emission were present, unless the dust attenuation on the gas is larger than in the stellar component. 

We tested the hybrid Pop III stellar templates from \citet{Rusta2025} for the fitting of {\it Pseudo-LRD-NOM}'s SED. We found that the resulting best-fit model yielded a reduced $\chi^2$ which is $>2\sigma$ higher than the value obtained for the SED fitting with the stellar models considered in \S\ref{sec:sed}. Thus, the presence of Pop~III stars does not appear to be significant in our source. A wider investigation of sources like {\it Pseudo-LRD-NOM}, and dwarf galaxies in general at high $z$, is necessary to find sources with a significant presence of first stellar generations.

\section{Conclusions}
\label{sec:concl}

We presented a detailed study of {\it Pseudo-LRD-NOM}, a compact, low-mass galaxy at $z=5.96$ which appears to be at the early stages of star-formation activity.  This source would fail the LRD classification because of its red rest-frame optical colour being contaminated by the presence of a prominent H$\alpha$ line.  It would comply instead with the LBD selection criteria recently proposed by \citet{Brazzini2026}.

We inferred the presence of an active, massive black hole in {\it Pseudo-LRD-NOM} from H$\alpha$ having a very broad component, whose large FWHM cannot be explained by stellar kinematics. We argued that the study of pseudo-LRDs has so far been largely neglected in the literature,  still it could be important to obtain a complete AGN census at high redshifts. 

Moreover, {\it Pseudo-LRD-NOM} provides evidence for an active black hole being embedded in a high-density, dusty starburst which is still at the early stages of its chemical enrichment. {\it Pseudo-LRD-NOM} is likely the precursor to a real LRD, with the massive central black-hole already in place, but without yet having the ultra-dense cocoon that would produce a prominent Balmer break in its spectrum. The study of more pseudo-LRDs residing in low-mass galaxies could be a key route to investigate the connection between massive black-hole growth and nuclear star clusters, as well as the LRD formation mechanisms. In addition, integral field spectroscopic studies with medium/high spectral resolution, with JWST and future 30-40~m-class ground-based telescopes, will be necessary to reveal in detail the dynamics and other physical processes at play in these sources.

\acknowledgments

We are grateful to Daniela Calzetti, Andrea Ferrara, Kohei Inayoshi, Roberto Maiolino, Gabriele Pezzulli, Fengwu Sun, and Zu Yan for useful discussions; Jos\'e M. Diego for providing us the most updated magnification value for our source; Raphael Hviding for making his Python package \textsc{unite} public and instructing us for its use; and Elka Rusta and Stefania Salvadori for providing us their galaxy SED models with hybrid Pop III stellar populations. We also thank the anonymous referee for a useful and constructive report.

This work is based on observations made with the NASA/ESA/CSA JWST. The data were obtained from the Mikulski Archive for Space Telescopes at the Space Telescope Science Institute, which is operated by the Association of Universities for Research in Astronomy, Inc., under NASA contract NAS 5-03127 for JWST. These observations are associated with the JWST GTO program PID 1208.  The specific observations analyzed here can be accessed via \dataset[doi: 10.17909/pyr9-5r43]{https://doi.org/10.17909/pyr9-5r43}. Some of the analyzed data products were retrieved from the Dawn JWST Archive (DJA). DJA is an initiative of the Cosmic Dawn Center (DAWN), which is funded by the Danish National Research Foundation under grant DNRF140. Also based on observations made with the NASA/ESA Hubble Space Telescope obtained from the Space Telescope Science Institute, which is operated by the Association of Universities for Research in Astronomy, Inc., under NASA contract NAS 526555.

KIC and RNC acknowledge funding from the Dutch Research Council
(NWO) through the award of the Vici Grant VI.C.212.036.

\vspace{5mm}
\facilities{{\sl JWST, HST}}.

\software{\textsc{Astropy} \citep{astropy_collaboration_astropy_2022}, 
\textsc{Cigale} \citep{boquien_cigale_2019},
\textsc{Cloudy} \citep{ferland_2013,ferland_2017},
\textsc{Msaexp}
\citep{brammer_msaexp_2023},
          \textsc{NumPy} \citep{harris_array_2020},
          \textsc{pandas} \citep{team_pandas-devpandas_2024},
          \textsc{Photutils} \citep{bradley_photutils_2016}, 
          \textsc{Topcat} \citep{taylor_topcat_2022},
          \textsc{unite} \citep{Hviding2025}.
          }
\bibliography{references}{}

@article{reddy_mosdef_2018,
	title = {The {MOSDEF} {Survey}: {Significant} {Evolution} in the {Rest}-frame {Optical} {Emission} {Line} {Equivalent} {Widths} of {Star}-forming {Galaxies} at z = 1.4-3.8},
	volume = {869},
	issn = {0004-637X},
	shorttitle = {The {MOSDEF} {Survey}},
	url = {https://ui.adsabs.harvard.edu/abs/2018ApJ...869...92R},
	doi = {10.3847/1538-4357/aaed1e},
	urldate = {2025-03-26},
	journal = {The Astrophysical Journal},
	author = {Reddy, Naveen A. and Shapley, Alice E. and Sanders, Ryan L. and Kriek, Mariska and Coil, Alison L. and Shivaei, Irene and Freeman, William R. and Mobasher, Bahram and Siana, Brian and Azadi, Mojegan and Fetherolf, Tara and Fornasini, Francesca M. and Leung, Gene and Price, Sedona H. and Zick, Tom and Barro, Guillermo},
	month = dec,
	year = {2018},
	note = {},
	pages = {92},
}

@article{ferruit_near-infrared_2022,
	title = {The {Near}-{Infrared} {Spectrograph} ({NIRSpec}) on the {James} {Webb} {Space} {Telescope}. {II}. {Multi}-object spectroscopy ({MOS})},
	volume = {661},
	issn = {0004-6361},
	url = {https://ui.adsabs.harvard.edu/abs/2022A&A...661A..81F},
	doi = {10.1051/0004-6361/202142673},
	urldate = {2025-03-26},
	journal = {Astronomy and Astrophysics},
	author = {Ferruit, P. and Jakobsen, P. and Giardino, G. and Rawle, T. and Alves de Oliveira, C. and Arribas, S. and Beck, T. L. and Birkmann, S. and Böker, T. and Bunker, A. J. and Charlot, S. and de Marchi, G. and Franx, M. and Henry, A. and Karakla, D. and Kassin, S. A. and Kumari, N. and López-Caniego, M. and Lützgendorf, N. and Maiolino, R. and Manjavacas, E. and Marston, A. and Moseley, S. H. and Muzerolle, J. and Pirzkal, N. and Rauscher, B. and Rix, H. -W. and Sabbi, E. and Sirianni, M. and te Plate, M. and Valenti, J. and Willott, C. J. and Zeidler, P.},
	month = may,
	year = {2022},
	note = {},
	pages = {A81},
}

@article{navarro-carrera_constraints_2024,
	title = {Constraints on the {Faint} {End} of the {Galaxy} {Stellar} {Mass} {Function} at z ≃ 4–8 from {Deep} {JWST} {Data}},
	volume = {961},
	issn = {0004-637X},
	url = {https://ui.adsabs.harvard.edu/abs/2024ApJ...961..207N},
	doi = {10.3847/1538-4357/ad0df6},
	urldate = {2025-03-15},
	journal = {The Astrophysical Journal},
	author = {Navarro-Carrera, Rafael and Rinaldi, Pierluigi and Caputi, Karina I. and Iani, Edoardo and Kokorev, Vasily and van Mierlo, Sophie E.},
	month = feb,
	year = {2024},
	note = {Publisher: IOP
ADS Bibcode: 2024ApJ...961..207N},
	pages = {207},
}

@article{iani_midis_2024-1,
	title = {{MIDIS}: {JWST} {NIRCam} and {MIRI} {Unveil} the {Stellar} {Population} {Properties} of {Lyα} {Emitters} and {Lyman}-break {Galaxies} at z ≃ 3–7},
	volume = {963},
	issn = {0004-637X},
	shorttitle = {{MIDIS}},
	url = {https://ui.adsabs.harvard.edu/abs/2024ApJ...963...97I},
	doi = {10.3847/1538-4357/ad15f6},
	urldate = {2025-02-21},
	journal = {The Astrophysical Journal},
	author = {Iani, Edoardo and Caputi, Karina I. and Rinaldi, Pierluigi and Annunziatella, Marianna and Boogaard, Leindert A. and Östlin, Göran and Costantin, Luca and Gillman, Steven and Pérez-González, Pablo G. and Colina, Luis and Greve, Thomas R. and Wright, Gillian and Alonso-Herrero, Almudena and Álvarez-Márquez, Javier and Bik, Arjan and Bosman, Sarah E. I. and Crespo Gómez, Alejandro and Eckart, Andreas and Hjorth, Jens and Jermann, Iris and Labiano, Alvaro and Langeroodi, Danial and Melinder, Jens and Moutard, Thibaud and Peißker, Florian and Pye, John P. and Tikkanen, Tuomo V. and van der Werf, Paul P. and Walter, Fabian and Henning, Thomas K. and Lagage, Pierre-Olivier and van Dishoeck, Ewine F.},
	month = mar,
	year = {2024},
	note = {Publisher: IOP
ADS Bibcode: 2024ApJ...963...97I},
	pages = {97},
}

@article{boquien_cigale_2019,
	title = {{CIGALE}: a python {Code} {Investigating} {GALaxy} {Emission}},
	volume = {622},
	issn = {0004-6361},
	shorttitle = {{CIGALE}},
	url = {https://ui.adsabs.harvard.edu/abs/2019A&A...622A.103B},
	doi = {10.1051/0004-6361/201834156},
	urldate = {2025-02-10},
	journal = {Astronomy and Astrophysics},
	author = {Boquien, M. and Burgarella, D. and Roehlly, Y. and Buat, V. and Ciesla, L. and Corre, D. and Inoue, A. K. and Salas, H.},
	month = feb,
	year = {2019},
	note = {},
	pages = {A103},
}

@article{stalevski_dust_2016,
	title = {The dust covering factor in active galactic nuclei},
	volume = {458},
	issn = {0035-8711},
	url = {},
	doi = {10.1093/mnras/stw444},
	urldate = {2025-02-10},
	journal = {Monthly Notices of the Royal Astronomical Society},
	author = {Stalevski, Marko and Ricci, Claudio and Ueda, Yoshihiro and Lira, Paulina and Fritz, Jacopo and Baes, Maarten},
	month = may,
	year = {2016},
	note = {},
	pages = {2288--2302},
}

@article{chabrier_galactic_2003,
	title = {Galactic {Stellar} and {Substellar} {Initial} {Mass} {Function}},
	volume = {115},
	issn = {0004-6280},
	url = {https://ui.adsabs.harvard.edu/abs/2003PASP..115..763C},
	doi = {10.1086/376392},
	urldate = {2025-02-10},
	journal = {Publications of the Astronomical Society of the Pacific},
	author = {Chabrier, Gilles},
	month = jul,
	year = {2003},
	note = {},
	pages = {763--795},
}

@article{caputi_alma_2021,
	title = {{ALMA} {Lensing} {Cluster} {Survey}: {An} {ALMA} {Galaxy} {Signposting} a {MUSE} {Galaxy} {Group} at z = 4.3 {Behind} "{El} {Gordo}"},
	volume = {908},
	issn = {0004-637X},
	shorttitle = {{ALMA} {Lensing} {Cluster} {Survey}},
	url = {https://ui.adsabs.harvard.edu/abs/2021ApJ...908..146C},
	doi = {10.3847/1538-4357/abd4d0},
	urldate = {2025-02-07},
	journal = {The Astrophysical Journal},
	author = {Caputi, K. I. and Caminha, G. B. and Fujimoto, S. and Kohno, K. and Sun, F. and Egami, E. and Deshmukh, S. and Tang, F. and Ao, Y. and Bradley, L. and Coe, D. and Espada, D. and Grillo, C. and Hatsukade, B. and Knudsen, K. K. and Lee, M. M. and Magdis, G. E. and Morokuma-Matsui, K. and Oesch, P. and Ouchi, M. and Rosati, P. and Umehata, H. and Valentino, F. and Vanzella, E. and Wang, W. -H. and Wu, J. F. and Zitrin, A.},
	month = feb,
	year = {2021},
	note = {},
	pages = {146},
}

@article{caputi_star_2017,
	title = {Star {Formation} in {Galaxies} at z ∼ 4-5 from the {SMUVS} {Survey}: {A} {Clear} {Starburst}/{Main}-sequence {Bimodality} for {Hα} {Emitters} on the {SFR}-{M}* {Plane}},
	volume = {849},
	issn = {0004-637X},
	shorttitle = {Star {Formation} in {Galaxies} at z ∼ 4-5 from the {SMUVS} {Survey}},
	url = {https://ui.adsabs.harvard.edu/abs/2017ApJ...849...45C},
	doi = {10.3847/1538-4357/aa901e},
	urldate = {2025-02-07},
	journal = {The Astrophysical Journal},
	author = {Caputi, K. I. and Deshmukh, S. and Ashby, M. L. N. and Cowley, W. I. and Bisigello, L. and Fazio, G. G. and Fynbo, J. P. U. and Le Fèvre, O. and Milvang-Jensen, B. and Ilbert, O.},
	month = nov,
	year = {2017},
	note = {},
	pages = {45},
}

@article{eldridge_binary_2017,
	title = {Binary {Population} and {Spectral} {Synthesis} {Version} 2.1: {Construction}, {Observational} {Verification}, and {New} {Results}},
	volume = {34},
	issn = {1323-3580},
	shorttitle = {Binary {Population} and {Spectral} {Synthesis} {Version} 2.1},
	url = {https://ui.adsabs.harvard.edu/abs/2017PASA...34...58E},
	doi = {10.1017/pasa.2017.51},
	urldate = {2025-02-06},
	journal = {Publications of the Astronomical Society of Australia},
	author = {Eldridge, J. J. and Stanway, E. R. and Xiao, L. and McClelland, L. A. S. and Taylor, G. and Ng, M. and Greis, S. M. L. and Bray, J. C.},
	month = nov,
	year = {2017},
	note = {ADS Bibcode: 2017PASA...34...58E},
	pages = {e058},
}

@article{ferland_2017,
	title = {The 2017 {Release} {Cloudy}},
	volume = {53},
	issn = {0185-1101},
	url = {https://ui.adsabs.harvard.edu/abs/2017RMxAA..53..385F},
	doi = {10.48550/arXiv.1705.10877},
	urldate = {2025-02-06},
	journal = {Revista Mexicana de Astronomia y Astrofisica},
	author = {Ferland, G. J. and Chatzikos, M. and Guzmán, F. and Lykins, M. L. and van Hoof, P. A. M. and Williams, R. J. R. and Abel, N. P. and Badnell, N. R. and Keenan, F. P. and Porter, R. L. and Stancil, P. C.},
	month = oct,
	year = {2017},
	note = {},
	pages = {385--438},
}

@article{harikane_jwstnirspec_2023,
	title = {A {JWST}/{NIRSpec} {First} {Census} of {Broad}-line {AGNs} at z = 4-7: {Detection} of 10 {Faint} {AGNs} with {M} {BH} 106-108 {M} ⊙ and {Their} {Host} {Galaxy} {Properties}},
	volume = {959},
	issn = {0004-637X},
	shorttitle = {A {JWST}/{NIRSpec} {First} {Census} of {Broad}-line {AGNs} at z = 4-7},
	url = {https://ui.adsabs.harvard.edu/abs/2023ApJ...959...39H},
	doi = {10.3847/1538-4357/ad029e},
	urldate = {2025-01-27},
	journal = {The Astrophysical Journal},
	author = {Harikane, Yuichi and Zhang, Yechi and Nakajima, Kimihiko and Ouchi, Masami and Isobe, Yuki and Ono, Yoshiaki and Hatano, Shun and Xu, Yi and Umeda, Hiroya},
	month = dec,
	year = {2023},
	note = {},
	pages = {39},
}

@article{asplund_chemical_2009,
	title = {The {Chemical} {Composition} of the {Sun}},
	volume = {47},
	issn = {0066-4146},
    url = {},
	doi = {10.1146/annurev.astro.46.060407.145222},
	urldate = {2025-01-25},
	journal = {Annual Review of Astronomy and Astrophysics},
	author = {Asplund, Martin and Grevesse, Nicolas and Sauval, A. Jacques and Scott, Pat},
	month = sep,
	year = {2009},
	note = {},
	pages = {481--522},
}

@book{osterbrock_astrophysics_2006,
	title = {Astrophysics of gaseous nebulae and active galactic nuclei},
	url = {https://ui.adsabs.harvard.edu/abs/2006agna.book.....O},
	urldate = {2025-01-19},
	author = {Osterbrock, Donald E. and Ferland, Gary J.},
	month = jan,
	year = {2006},
	note = {},
}

@ARTICLE{Rinaldi2025_notdot,
       author = {{Rinaldi}, P. and {Bonaventura}, N. and {Rieke}, G.~H. and {Alberts}, S. and {Caputi}, K.~I. and {Baker}, W.~M. and {Baum}, S. and {Bhatawdekar}, R. and {Bunker}, A.~J. and {Carniani}, S. and {Curtis-Lake}, E. and {D'Eugenio}, F. and {Egami}, E. and {Ji}, Z. and {Johnson}, B.~D. and {Hainline}, K. and {Helton}, J.~M. and {Lin}, X. and {Lyu}, J. and {Ma}, Z. and {Maiolino}, R. and {P{\'e}rez-Gonz{\'a}lez}, P.~G. and {Rieke}, M. and {Robertson}, B.~E. and {Shivaei}, I. and {Stone}, M. and {Sun}, Y. and {Tacchella}, S. and {{\"U}bler}, H. and {Williams}, C.~C. and {Willmer}, C.~N.~A. and {Willott}, C. and {Zhang}, J. and {Zhu}, Y.},
        title = "{Not Just a Dot: The Complex UV Morphology and Underlying Properties of Little Red Dots}",
      journal = {\apj},
         year = 2025,
        month = oct,
       volume = {992},
       number = {1},
          eid = {71},
        pages = {71},
          doi = {10.3847/1538-4357/adfa10},
archivePrefix = {arXiv},
       eprint = {2411.14383},
 primaryClass = {astro-ph.GA},
       adsurl = {https://ui.adsabs.harvard.edu/abs/2025ApJ...992...71R}
}

@article{gardner_james_2023,
	title = {The {James} {Webb} {Space} {Telescope} {Mission}},
	volume = {135},
	issn = {0004-6280},
	url = {https://ui.adsabs.harvard.edu/abs/2023PASP..135f8001G},
	doi = {10.1088/1538-3873/acd1b5},
	urldate = {2025-01-15},
	journal = {Publications of the Astronomical Society of the Pacific},
	author = {Gardner, Jonathan P. and Mather, John C. and Abbott, Randy and Abell, James S. and Abernathy, Mark and Abney, Faith E. and Abraham, John G. and Abraham, Roberto and Abul-Huda, Yasin M. and Acton, Scott and Adams, Cynthia K. and Adams, Evan and Adler, David S. and Adriaensen, Maarten and Aguilar, Jonathan Albert and Ahmed, Mansoor and Ahmed, Nasif S. and Ahmed, Tanjira and Albat, Rüdeger and Albert, Loïc and Alberts, Stacey and Aldridge, David and Allen, Mary Marsha and Allen, Shaune S. and Altenburg, Martin and Altunc, Serhat and Alvarez, Jose Lorenzo and Álvarez-Márquez, Javier and Alves de Oliveira, Catarina and Ambrose, Leslie L. and Anandakrishnan, Satya M. and Andersen, Gregory C. and Anderson, Harry James and Anderson, Jay and Anderson, Kristen and Anderson, Sara M. and Aprea, Julio and Archer, Benita J. and Arenberg, Jonathan W. and Argyriou, Ioannis and Arribas, Santiago and Artigau, Étienne and Arvai, Amanda Rose and Atcheson, Paul and Atkinson, Charles B. and Averbukh, Jesse and Aymergen, Cagatay and Bacinski, John J. and Baggett, Wayne E. and Bagnasco, Giorgio and Baker, Lynn L. and Balzano, Vicki Ann and Banks, Kimberly A. and Baran, David A. and Barker, Elizabeth A. and Barrett, Larry K. and Barringer, Bruce O. and Barto, Allison and Bast, William and Baudoz, Pierre and Baum, Stefi and Beatty, Thomas G. and Beaulieu, Mathilde and Bechtold, Kathryn and Beck, Tracy and Beddard, Megan M. and Beichman, Charles and Bellagama, Larry and Bely, Pierre and Berger, Timothy W. and Bergeron, Louis E. and Bernier, Antoine-Darveau and Bertch, Maria D. and Beskow, Charlotte and Betz, Laura E. and Biagetti, Carl P. and Birkmann, Stephan and Bjorklund, Kurt F. and Blackwood, James D. and Blazek, Ronald Paul and Blossfeld, Stephen and Bluth, Marcel and Boccaletti, Anthony and Boegner, Jr., Martin E. and Bohlin, Ralph C. and Boia, John Joseph and Böker, Torsten and Bonaventura, N. and Bond, Nicholas A. and Bosley, Kari Ann and Boucarut, Rene A. and Bouchet, Patrice and Bouwman, Jeroen and Bower, Gary and Bowers, Ariel S. and Bowers, Charles W. and Boyce, Leslye A. and Boyer, Christine T. and Boyer, Martha L. and Boyer, Michael and Boyer, Robert and Bradley, Larry D. and Brady, Gregory R. and Brandl, Bernhard R. and Brannen, Judith L. and Breda, David and Bremmer, Harold G. and Brennan, David and Bresnahan, Pamela A. and Bright, Stacey N. and Broiles, Brian J. and Bromenschenkel, Asa and Brooks, Brian H. and Brooks, Keira J. and Brown, Bob and Brown, Bruce and Brown, Thomas M. and Bruce, Barry W. and Bryson, Jonathan G. and Bujanda, Edwin D. and Bullock, Blake M. and Bunker, A. J. and Bureo, Rafael and Burt, Irving J. and Bush, James Aaron and Bushouse, Howard A. and Bussman, Marie C. and Cabaud, Olivier and Cale, Steven and Calhoon, Charles D. and Calvani, Humberto and Canipe, Alicia M. and Caputo, Francis M. and Cara, Mihai and Carey, Larkin and Case, Michael Eli and Cesari, Thaddeus and Cetorelli, Lee D. and Chance, Don R. and Chandler, Lynn and Chaney, Dave and Chapman, George N. and Charlot, S. and Chayer, Pierre and Cheezum, Jeffrey I. and Chen, Bin and Chen, Christine H. and Cherinka, Brian and Chichester, Sarah C. and Chilton, Zachary S. and Chittiraibalan, Dharini and Clampin, Mark and Clark, Charles R. and Clark, Kerry W. and Clark, Stephanie M. and Claybrooks, Edward E. and Cleveland, Keith A. and Cohen, Andrew L. and Cohen, Lester M. and Colón, Knicole D. and Coleman, Benee L. and Colina, Luis and Comber, Brian J. and Comeau, Thomas M. and Comer, Thomas and Conde Reis, Alain and Connolly, Dennis C. and Conroy, Kyle E. and Contos, Adam R. and Contreras, James and Cook, Neil J. and Cooper, James L. and Cooper, Rachel Aviva and Correia, Michael F. and Correnti, Matteo and Cossou, Christophe and Costanza, Brian F. and Coulais, Alain and Cox, Colin R. and Coyle, Ray T. and Cracraft, Misty M. and Crew, Keith A. and Curtis, Gary J. and Cusveller, Bianca and Da Costa Maciel, Cleyciane and Dailey, Christopher T. and Daugeron, Frédéric and Davidson, Greg S. and Davies, James E. and Davis, Katherine Anne and Davis, Michael S. and Day, Ratna and de Chambure, Daniel and de Jong, Pauline and De Marchi, Guido and Dean, Bruce H. and Decker, John E. and Delisa, Amy S. and Dell, Lawrence C. and Dellagatta, Gail},
	month = jun,
	year = {2023},
	note = {},
	pages = {068001},
}

@article{greene_intermediate-mass_2020,
	title = {Intermediate-{Mass} {Black} {Holes}},
	volume = {58},
	issn = {0066-4146},
	url = {https://ui.adsabs.harvard.edu/abs/2020ARA&A..58..257G},
	doi = {10.1146/annurev-astro-032620-021835},
	urldate = {2024-11-20},
	journal = {Annual Review of Astronomy and Astrophysics},
	author = {Greene, Jenny E. and Strader, Jay and Ho, Luis C.},
	month = aug,
	year = {2020},
	note = {},
	pages = {257--312},
}

@article{greene_uncover_2024,
	title = {{UNCOVER} {Spectroscopy} {Confirms} the {Surprising} {Ubiquity} of {Active} {Galactic} {Nuclei} in {Red} {Sources} at z {\textgreater} 5},
	volume = {964},
	issn = {0004-637X},
	url = {https://ui.adsabs.harvard.edu/abs/2024ApJ...964...39G},
	doi = {10.3847/1538-4357/ad1e5f},
	urldate = {2024-11-17},
	journal = {The Astrophysical Journal},
	author = {Greene, Jenny E. and Labbe, Ivo and Goulding, Andy D. and Furtak, Lukas J. and Chemerynska, Iryna and Kokorev, Vasily and Dayal, Pratika and Volonteri, Marta and Williams, Christina C. and Wang, Bingjie and Setton, David J. and Burgasser, Adam J. and Bezanson, Rachel and Atek, Hakim and Brammer, Gabriel and Cutler, Sam E. and Feldmann, Robert and Fujimoto, Seiji and Glazebrook, Karl and de Graaff, Anna and Khullar, Gourav and Leja, Joel and Marchesini, Danilo and Maseda, Michael V. and Matthee, Jorryt and Miller, Tim B. and Naidu, Rohan P. and Nanayakkara, Themiya and Oesch, Pascal A. and Pan, Richard and Papovich, Casey and Price, Sedona H. and van Dokkum, Pieter and Weaver, John R. and Whitaker, Katherine E. and Zitrin, Adi},
	month = mar,
	year = {2024},
	note = {},
	pages = {39},
}

@misc{team_pandas-devpandas_2024,
	title = {pandas-dev/pandas: {Pandas}},
	shorttitle = {pandas-dev/pandas},
	url = {https://zenodo.org/records/13819579},
	urldate = {2024-11-17},
	publisher = {Zenodo},
	author = {team, The pandas development},
	month = sep,
	year = {2024},
	doi = {10.5281/zenodo.13819579},
}

@article{brammer_msaexp_2023,
	title = {msaexp: {NIRSpec} analyis tools},
	shorttitle = {msaexp},
	url = {https://ui.adsabs.harvard.edu/abs/2022zndo...7299500B},
	doi = {10.5281/zenodo.7299500},
	urldate = {2024-11-17},
	journal = {Zenodo},
	author = {Brammer, Gabriel},
	month = sep,
	year = {2023},
	note = {},
}

@article{taylor_topcat_2022,
	title = {{TOPCAT} {Visualisation} {Over} the {Web}},
	volume = {532},
	url = {https://ui.adsabs.harvard.edu/abs/2022ASPC..532....3T},
	urldate = {2024-11-17},
	author = {Taylor, Mark},
	month = jul,
	year = {2022},
	note = {Conference Name: Astronomical Data Analysis Software and Systems XXX},
	pages = {3},
}

@article{bradley_photutils_2016,
	title = {Photutils: {Photometry} tools},
	shorttitle = {Photutils},
	url = {https://ui.adsabs.harvard.edu/abs/2016ascl.soft09011B},
	urldate = {2024-11-17},
	journal = {Astrophysics Source Code Library},
	author = {Bradley, Larry and Sipocz, Brigitta and Robitaille, Thomas and Tollerud, Erik and Deil, Christoph and Vinícius, Zè and Barbary, Kyle and Günther, Hans Moritz and Bostroem, Azalee and Droettboom, Michael and Bray, Erik and Bratholm, Lars Andersen and Pickering, T. E. and Craig, Matt and Pascual, Sergio and Greco, Johnny and Donath, Axel and Kerzendorf, Wolfgang and Littlefair, Stuart and Barentsen, Geert and D'Eugenio, Francesco and Weaver, Benjamin Alan},
	month = sep,
	year = {2016},
	note = {},
	pages = {ascl:1609.011},
}

@article{astropy_collaboration_astropy_2022,
	title = {The {Astropy} {Project}: {Sustaining} and {Growing} a {Community}-oriented {Open}-source {Project} and the {Latest} {Major} {Release} (v5.0) of the {Core} {Package}},
	volume = {935},
	issn = {0004-637X},
	shorttitle = {The {Astropy} {Project}},
	url = {https://ui.adsabs.harvard.edu/abs/2022ApJ...935..167A},
	doi = {10.3847/1538-4357/ac7c74},
	urldate = {2024-11-17},
	journal = {The Astrophysical Journal},
	author = {{Astropy Collaboration} and Price-Whelan, Adrian M. and Lim, Pey Lian and Earl, Nicholas and Starkman, Nathaniel and Bradley, Larry and Shupe, David L. and Patil, Aarya A. and Corrales, Lia and Brasseur, C. E. and Nöthe, Maximilian and Donath, Axel and Tollerud, Erik and Morris, Brett M. and Ginsburg, Adam and Vaher, Eero and Weaver, Benjamin A. and Tocknell, James and Jamieson, William and van Kerkwijk, Marten H. and Robitaille, Thomas P. and Merry, Bruce and Bachetti, Matteo and Günther, H. Moritz and Aldcroft, Thomas L. and Alvarado-Montes, Jaime A. and Archibald, Anne M. and Bódi, Attila and Bapat, Shreyas and Barentsen, Geert and Bazán, Juanjo and Biswas, Manish and Boquien, Médéric and Burke, D. J. and Cara, Daria and Cara, Mihai and Conroy, Kyle E. and Conseil, Simon and Craig, Matthew W. and Cross, Robert M. and Cruz, Kelle L. and D'Eugenio, Francesco and Dencheva, Nadia and Devillepoix, Hadrien A. R. and Dietrich, Jörg P. and Eigenbrot, Arthur Davis and Erben, Thomas and Ferreira, Leonardo and Foreman-Mackey, Daniel and Fox, Ryan and Freij, Nabil and Garg, Suyog and Geda, Robel and Glattly, Lauren and Gondhalekar, Yash and Gordon, Karl D. and Grant, David and Greenfield, Perry and Groener, Austen M. and Guest, Steve and Gurovich, Sebastian and Handberg, Rasmus and Hart, Akeem and Hatfield-Dodds, Zac and Homeier, Derek and Hosseinzadeh, Griffin and Jenness, Tim and Jones, Craig K. and Joseph, Prajwel and Kalmbach, J. Bryce and Karamehmetoglu, Emir and Kałuszyński, Mikołaj and Kelley, Michael S. P. and Kern, Nicholas and Kerzendorf, Wolfgang E. and Koch, Eric W. and Kulumani, Shankar and Lee, Antony and Ly, Chun and Ma, Zhiyuan and MacBride, Conor and Maljaars, Jakob M. and Muna, Demitri and Murphy, N. A. and Norman, Henrik and O'Steen, Richard and Oman, Kyle A. and Pacifici, Camilla and Pascual, Sergio and Pascual-Granado, J. and Patil, Rohit R. and Perren, Gabriel I. and Pickering, Timothy E. and Rastogi, Tanuj and Roulston, Benjamin R. and Ryan, Daniel F. and Rykoff, Eli S. and Sabater, Jose and Sakurikar, Parikshit and Salgado, Jesús and Sanghi, Aniket and Saunders, Nicholas and Savchenko, Volodymyr and Schwardt, Ludwig and Seifert-Eckert, Michael and Shih, Albert Y. and Jain, Anany Shrey and Shukla, Gyanendra and Sick, Jonathan and Simpson, Chris and Singanamalla, Sudheesh and Singer, Leo P. and Singhal, Jaladh and Sinha, Manodeep and Sipőcz, Brigitta M. and Spitler, Lee R. and Stansby, David and Streicher, Ole and Šumak, Jani and Swinbank, John D. and Taranu, Dan S. and Tewary, Nikita and Tremblay, Grant R. and de Val-Borro, Miguel and Van Kooten, Samuel J. and Vasović, Zlatan and Verma, Shresth and de Miranda Cardoso, José Vinícius and Williams, Peter K. G. and Wilson, Tom J. and Winkel, Benjamin and Wood-Vasey, W. M. and Xue, Rui and Yoachim, Peter and Zhang, Chen and Zonca, Andrea and {Astropy Project Contributors}},
	month = aug,
	year = {2022},
	note = {},
	pages = {167},
}

@article{harris_array_2020,
	title = {Array programming with {NumPy}},
	volume = {585},
	issn = {0028-0836},
	url = {https://ui.adsabs.harvard.edu/abs/2020Natur.585..357H},
	doi = {10.1038/s41586-020-2649-2},
	urldate = {2024-11-17},
	journal = {Nature},
	author = {Harris, Charles R. and Millman, K. Jarrod and van der Walt, Stéfan J. and Gommers, Ralf and Virtanen, Pauli and Cournapeau, David and Wieser, Eric and Taylor, Julian and Berg, Sebastian and Smith, Nathaniel J. and Kern, Robert and Picus, Matti and Hoyer, Stephan and van Kerkwijk, Marten H. and Brett, Matthew and Haldane, Allan and del Río, Jaime Fernández and Wiebe, Mark and Peterson, Pearu and Gérard-Marchant, Pierre and Sheppard, Kevin and Reddy, Tyler and Weckesser, Warren and Abbasi, Hameer and Gohlke, Christoph and Oliphant, Travis E.},
	month = sep,
	year = {2020},
	note = {},
	pages = {357--362},
}

@article{ubler_ga-nifs_2023,
	title = {{GA}-{NIFS}: {A} massive black hole in a low-metallicity {AGN} at z ∼ 5.55 revealed by {JWST}/{NIRSpec} {IFS}},
	volume = {677},
	issn = {0004-6361},
	shorttitle = {{GA}-{NIFS}},
	url = {https://ui.adsabs.harvard.edu/abs/2023A&A...677A.145U/abstract},
	doi = {10.1051/0004-6361/202346137},
	language = {en},
	urldate = {2024-11-17},
	journal = {Astronomy and Astrophysics},
	author = {\"Ubler, Hannah and Maiolino, Roberto and Curtis-Lake, Emma and Pérez-González, Pablo G. and Curti, Mirko and Perna, Michele and Arribas, Santiago and Charlot, Stéphane and Marshall, Madeline A. and D'Eugenio, Francesco and Scholtz, Jan and Bunker, Andrew and Carniani, Stefano and Ferruit, Pierre and Jakobsen, Peter and Rix, Hans-Walter and Rodríguez Del Pino, Bruno and Willott, Chris J. and Boeker, Torsten and Cresci, Giovanni and Jones, Gareth C. and Kumari, Nimisha and Rawle, Tim},
	month = sep,
	year = {2023},
	pages = {A145},
}

@ARTICLE{Maiolino2025_Chandra,
       author = {{Maiolino}, Roberto and {Risaliti}, Guido and {Signorini}, Matilde and {Trefoloni}, Bartolomeo and {Juod{\v{z}}balis}, Ignas and {Scholtz}, Jan and {{\"U}bler}, Hannah and {D'Eugenio}, Francesco and {Carniani}, Stefano and {Fabian}, Andy and {Ji}, Xihan and {Mazzolari}, Giovanni and {Bertola}, Elena and {Brusa}, Marcella and {Bunker}, Andrew J. and {Charlot}, Stephane and {Comastri}, Andrea and {Cresci}, Giovanni and {DeCoursey}, Christa Noel and {Egami}, Eiichi and {Fiore}, Fabrizio and {Gilli}, Roberto and {Perna}, Michele and {Tacchella}, Sandro and {Venturi}, Giacomo},
        title = "{JWST meets Chandra: a large population of Compton thick, feedback-free, and intrinsically X-ray weak AGN, with a sprinkle of SNe}",
      journal = {\mnras},
         year = 2025,
        month = apr,
       volume = {538},
       number = {3},
        pages = {1921-1943},
          doi = {10.1093/mnras/staf359},
archivePrefix = {arXiv},
       eprint = {2405.00504},
 primaryClass = {astro-ph.GA},
       adsurl = {https://ui.adsabs.harvard.edu/abs/2025MNRAS.538.1921M}
}

@article{magorrian_demography_1998,
	title = {The {Demography} of {Massive} {Dark} {Objects} in {Galaxy} {Centers}},
	volume = {115},
	issn = {0004-6256},
	url = {https://ui.adsabs.harvard.edu/abs/1998AJ....115.2285M},
	doi = {10.1086/300353},
	urldate = {2024-09-15},
	journal = {The Astronomical Journal},
	author = {Magorrian, John and Tremaine, Scott and Richstone, Douglas and Bender, Ralf and Bower, Gary and Dressler, Alan and Faber, S. M. and Gebhardt, Karl and Green, Richard and Grillmair, Carl and Kormendy, John and Lauer, Tod},
	month = jun,
	year = {1998},
	note = {},
	pages = {2285--2305},
}

@article{pacucci_jwst_2023,
	title = {{JWST} {CEERS} and {JADES} {Active} {Galaxies} at z = 4-7 {Violate} the {Local} {M} •-{M} ⋆ {Relation} at {\textgreater}3σ: {Implications} for {Low}-mass {Black} {Holes} and {Seeding} {Models}},
	volume = {957},
	issn = {0004-637X},
	shorttitle = {{JWST} {CEERS} and {JADES} {Active} {Galaxies} at z = 4-7 {Violate} the {Local} {M} •-{M} ⋆ {Relation} at {\textgreater}3σ},
	url = {https://ui.adsabs.harvard.edu/abs/2023ApJ...957L...3P},
	doi = {10.3847/2041-8213/ad0158},
	urldate = {2024-09-15},
	journal = {The Astrophysical Journal},
	author = {Pacucci, Fabio and Nguyen, Bao and Carniani, Stefano and Maiolino, Roberto and Fan, Xiaohui},
	month = nov,
	year = {2023},
	note = {},
	pages = {L3},
}

@article{nakajima_diagnostics_2022,
	title = {Diagnostics for {PopIII} galaxies and direct collapse black holes in the early universe},
	volume = {513},
	issn = {0035-8711},
	url = {https://ui.adsabs.harvard.edu/abs/2022MNRAS.513.5134N},
	doi = {10.1093/mnras/stac1242},
	urldate = {2024-09-15},
	journal = {Monthly Notices of the Royal Astronomical Society},
	author = {Nakajima, K. and Maiolino, R.},
	month = jul,
	year = {2022},
	note = {},
	pages = {5134--5147},
}

@article{calzetti_dust_2000,
	title = {The {Dust} {Content} and {Opacity} of {Actively} {Star}-forming {Galaxies}},
	volume = {533},
	issn = {0004-637X},
	url = {https://ui.adsabs.harvard.edu/abs/2000ApJ...533..682C},
	doi = {10.1086/308692},
	urldate = {2024-09-14},
	journal = {The Astrophysical Journal},
	author = {Calzetti, Daniela and Armus, Lee and Bohlin, Ralph C. and Kinney, Anne L. and Koornneef, Jan and Storchi-Bergmann, Thaisa},
	month = apr,
	year = {2000},
	note = {},
	pages = {682--695},
}

@article{ferland_2013,
	title = {The 2013 {Release} of {Cloudy}},
	volume = {49},
	issn = {0185-1101},
	url = {https://ui.adsabs.harvard.edu/abs/2013RMxAA..49..137F},
	doi = {10.48550/arXiv.1302.4485},
	urldate = {2024-09-14},
	journal = {Revista Mexicana de Astronomia y Astrofisica},
	author = {Ferland, G. J. and Porter, R. L. and van Hoof, P. A. M. and Williams, R. J. R. and Abel, N. P. and Lykins, M. L. and Shaw, G. and Henney, W. J. and Stancil, P. C.},
	month = apr,
	year = {2013},
	note = {},
	pages = {137--163},
}

@article{bruzual_stellar_2003,
	title = {Stellar population synthesis at the resolution of 2003},
	volume = {344},
	issn = {0035-8711},
	url = {https://ui.adsabs.harvard.edu/abs/2003MNRAS.344.1000B},
	doi = {10.1046/j.1365-8711.2003.06897.x},
	urldate = {2024-09-14},
	journal = {Monthly Notices of the Royal Astronomical Society},
	author = {Bruzual, G. and Charlot, S.},
	month = oct,
	year = {2003},
	note = {},
	pages = {1000--1028},
}

@ARTICLE{Labbe2025_uncover,
       author = {{Labb\'e}, Ivo and {Greene}, Jenny E. and {Bezanson}, Rachel and {Fujimoto}, Seiji and {Furtak}, Lukas J. and {Goulding}, Andy D. and {Matthee}, Jorryt and {Naidu}, Rohan P. and {Oesch}, Pascal A. and {Atek}, Hakim and {Brammer}, Gabriel and {Chemerynska}, Iryna and {Coe}, Dan and {Cutler}, Sam E. and {Dayal}, Pratika and {Feldmann}, Robert and {Franx}, Marijn and {Glazebrook}, Karl and {Leja}, Joel and {Maseda}, Michael and {Marchesini}, Danilo and {Nanayakkara}, Themiya and {Nelson}, Erica J. and {Pan}, Richard and {Papovich}, Casey and {Price}, Sedona H. and {Suess}, Katherine A. and {Wang}, Bingjie and {Weaver}, John R. and {Whitaker}, Katherine E. and {Williams}, Christina C. and {Zitrin}, Adi},
        title = "{UNCOVER: Candidate Red Active Galactic Nuclei at 3 < z < 7 with JWST and ALMA}",
      journal = {\apj},
         year = 2025,
        month = jan,
       volume = {978},
       number = {1},
          eid = {92},
        pages = {92},
          doi = {10.3847/1538-4357/ad3551},
archivePrefix = {arXiv},
       eprint = {2306.07320},
 primaryClass = {astro-ph.GA},
       adsurl = {https://ui.adsabs.harvard.edu/abs/2025ApJ...978...92L}
}

@ARTICLE{Onoue2019,
       author = {{Onoue}, Masafusa and {Kashikawa}, Nobunari and {Matsuoka}, Yoshiki and {Kato}, Nanako and {Izumi}, Takuma and {Nagao}, Tohru and {Strauss}, Michael A. and {Harikane}, Yuichi and {Imanishi}, Masatoshi and {Ito}, Kei and {Iwasawa}, Kazushi and {Kawaguchi}, Toshihiro and {Lee}, Chien-Hsiu and {Noboriguchi}, Akatoki and {Suh}, Hyewon and {Tanaka}, Masayuki and {Toba}, Yoshiki},
        title = "{Subaru High-z Exploration of Low-luminosity Quasars (SHELLQs). VI. Black Hole Mass Measurements of Six Quasars at 6.1 {\ensuremath{\leq}} z {\ensuremath{\leq}} 6.7}",
      journal = {\apj},
         year = 2019,
        month = aug,
       volume = {880},
       number = {2},
          eid = {77},
        pages = {77},
          doi = {10.3847/1538-4357/ab29e9},
archivePrefix = {arXiv},
       eprint = {1904.07278},
 primaryClass = {astro-ph.GA},
       adsurl = {https://ui.adsabs.harvard.edu/abs/2019ApJ...880...77O}
}

@article{perez-gonzalez_what_2024,
	title = {What {Is} the {Nature} of {Little} {Red} {Dots} and what {Is} {Not}, {MIRI} {SMILES} {Edition}},
	volume = {968},
	issn = {0004-637X},
	url = {https://ui.adsabs.harvard.edu/abs/2024ApJ...968....4P},
	doi = {10.3847/1538-4357/ad38bb},
	urldate = {2024-09-14},
	journal = {The Astrophysical Journal},
	author = {Pérez-González, Pablo G. and Barro, Guillermo and Rieke, George H. and Lyu, Jianwei and Rieke, Marcia and Alberts, Stacey and Williams, Christina C. and Hainline, Kevin and Sun, Fengwu and Puskás, Dávid and Annunziatella, Marianna and Baker, William M. and Bunker, Andrew J. and Egami, Eiichi and Ji, Zhiyuan and Johnson, Benjamin D. and Robertson, Brant and Rodríguez Del Pino, Bruno and Rujopakarn, Wiphu and Shivaei, Irene and Tacchella, Sandro and Willmer, Christopher N. A. and Willott, Chris},
	month = jun,
	year = {2024},
	note = {},
	pages = {4},
}

@article{kokorev_census_2024,
	title = {A {Census} of {Photometrically} {Selected} {Little} {Red} {Dots} at 4 {\textless} z {\textless} 9 in {JWST} {Blank} {Fields}},
	volume = {968},
	issn = {0004-637X},
	url = {https://ui.adsabs.harvard.edu/abs/2024ApJ...968...38K},
	doi = {10.3847/1538-4357/ad4265},
	urldate = {2024-09-14},
	journal = {The Astrophysical Journal},
	author = {Kokorev, Vasily and Caputi, Karina I. and Greene, Jenny E. and Dayal, Pratika and Trebitsch, Maxime and Cutler, Sam E. and Fujimoto, Seiji and Labbé, Ivo and Miller, Tim B. and Iani, Edoardo and Navarro-Carrera, Rafael and Rinaldi, Pierluigi},
	month = jun,
	year = {2024},
	note = {},
	pages = {38},
}

@ARTICLE{Kocevski2025,
       author = {{Kocevski}, Dale D. and {Finkelstein}, Steven L. and {Barro}, Guillermo and {Taylor}, Anthony J. and {Calabr{\`o}}, Antonello and {Laloux}, Brivael and {Buchner}, Johannes and {Trump}, Jonathan R. and {Leung}, Gene C.~K. and {Yang}, Guang and {Dickinson}, Mark and {P{\'e}rez-Gonz{\'a}lez}, Pablo G. and {Pacucci}, Fabio and {Inayoshi}, Kohei and {Somerville}, Rachel S. and {McGrath}, Elizabeth J. and {Akins}, Hollis B. and {Bagley}, Micaela B. and {Bowler}, Rebecca A.~A. and {Bisigello}, Laura and {Carnall}, Adam and {Casey}, Caitlin M. and {Cheng}, Yingjie and {Cleri}, Nikko J. and {Costantin}, Luca and {Cullen}, Fergus and {Davis}, Kelcey and {Donnan}, Callum T. and {Dunlop}, James S. and {Ellis}, Richard S. and {Ferguson}, Henry C. and {Fujimoto}, Seiji and {Fontana}, Adriano and {Giavalisco}, Mauro and {Grazian}, Andrea and {Grogin}, Norman A. and {Hathi}, Nimish P. and {Hirschmann}, Michaela and {Huertas-Company}, Marc and {Holwerda}, Benne W. and {Illingworth}, Garth and {Juneau}, St{\'e}phanie and {Kartaltepe}, Jeyhan S. and {Koekemoer}, Anton M. and {Li}, Wenxiu and {Lucas}, Ray A. and {Magee}, Dan and {Mason}, Charlotte and {McLeod}, Derek J. and {McLure}, Ross J. and {Napolitano}, Lorenzo and {Papovich}, Casey and {Pirzkal}, Nor and {Rodighiero}, Giulia and {Santini}, Paola and {Wilkins}, Stephen M. and {Yung}, L.~Y. Aaron},
        title = "{The Rise of Faint, Red Active Galactic Nuclei at z > 4: A Sample of Little Red Dots in the JWST Extragalactic Legacy Fields}",
      journal = {\apj},
         year = 2025,
        month = jun,
       volume = {986},
       number = {2},
          eid = {126},
        pages = {126},
          doi = {10.3847/1538-4357/adbc7d},
archivePrefix = {arXiv},
       eprint = {2404.03576},
 primaryClass = {astro-ph.GA},
       adsurl = {https://ui.adsabs.harvard.edu/abs/2025ApJ...986..126K}
}

@article{furtak_jwst_2023,
	title = {{JWST} {UNCOVER}: {Extremely} {Red} and {Compact} {Object} at z phot ≃ 7.6 {Triply} {Imaged} by {A2744}},
	volume = {952},
	issn = {0004-637X},
	shorttitle = {{JWST} {UNCOVER}},
	url = {https://ui.adsabs.harvard.edu/abs/2023ApJ...952..142F},
	doi = {10.3847/1538-4357/acdc9d},
	urldate = {2024-09-14},
	journal = {The Astrophysical Journal},
	author = {Furtak, Lukas J. and Zitrin, Adi and Plat, Adèle and Fujimoto, Seiji and Wang, Bingjie and Nelson, Erica J. and Labbé, Ivo and Bezanson, Rachel and Brammer, Gabriel B. and van Dokkum, Pieter and Endsley, Ryan and Glazebrook, Karl and Greene, Jenny E. and Leja, Joel and Price, Sedona H. and Smit, Renske and Stark, Daniel P. and Weaver, John R. and Whitaker, Katherine E. and Atek, Hakim and Chevallard, Jacopo and Curtis-Lake, Emma and Dayal, Pratika and Feltre, Anna and Franx, Marijn and Fudamoto, Yoshinobu and Marchesini, Danilo and Mowla, Lamiya A. and Pan, Richard and Suess, Katherine A. and Vidal-García, Alba and Williams, Christina C.},
	month = aug,
	year = {2023},
	note = {},
	pages = {142},
}

@article{caputi_midis_2024,
	title = {{MIDIS}: {The} {Relation} between {Strong} ({Hβ} + [{O} {III}]) {Emission}, {Star} {Formation}, and {Burstiness} around the {Epoch} of {Reionization}},
	volume = {969},
	issn = {0004-637X},
	shorttitle = {{MIDIS}},
	url = {https://ui.adsabs.harvard.edu/abs/2024ApJ...969..159C},
	doi = {10.3847/1538-4357/ad4eb2},
	urldate = {2024-09-14},
	journal = {The Astrophysical Journal},
	author = {Caputi, K. I. and Rinaldi, P. and Iani, E. and Pérez-González, P. G. and Östlin, G. and Colina, L. and Greve, T. R. and Nørgaard-Nielsen, H. U. and Wright, G. S. and Álvarez-Márquez, J. and Eckart, A. and Hjorth, J. and Labiano, A. and Le Fèvre, O. and Walter, F. and van der Werf, P. and Boogaard, L. and Costantin, L. and Crespo Gómez, A. and Gillman, S. and Jermann, I. and Langeroodi, D. and Melinder, J. and Peissker, F. and Güdel, M. and Henning, Th. and Lagage, P. O. and Ray, T. P.},
	month = jul,
	year = {2024},
	note = {},
	pages = {159},
}

@article{rinaldi_midis_2023,
	title = {{MIDIS}: {Strong} ({Hβ}+[{O} {III}]) and {Hα} {Emitters} at {Redshift} z ≃ 7-8 {Unveiled} with {JWST} {NIRCam} and {MIRI} {Imaging} in the {Hubble} {eXtreme} {Deep} {Field}},
	volume = {952},
	issn = {0004-637X},
	shorttitle = {{MIDIS}},
	url = {https://ui.adsabs.harvard.edu/abs/2023ApJ...952..143R},
	doi = {10.3847/1538-4357/acdc27},
	urldate = {2024-09-14},
	journal = {The Astrophysical Journal},
	author = {Rinaldi, P. and Caputi, K. I. and Costantin, L. and Gillman, S. and Iani, E. and Pérez-González, P. G. and Östlin, G. and Colina, L. and Greve, T. R. and Noorgard-Nielsen, H. U. and Wright, G. S. and Alonso-Herrero, A. and Álvarez-Márquez, J. and Eckart, A. and García-Marín, M. and Hjorth, J. and Ilbert, O. and Kendrew, S. and Labiano, A. and Le Fèvre, O. and Pye, J. and Tikkanen, T. and Walter, F. and van der Werf, P. and Ward, M. and Annunziatella, M. and Azzollini, R. and Bik, A. and Boogaard, L. and Bosman, S. E. I. and Crespo Gómez, A. and Jermann, I. and Langeroodi, D. and Melinder, J. and Meyer, R. A. and Moutard, T. and Peissker, F. and Topinka, M. and van Dishoeck, E. and Güdel, M. and Henning, Th. and Lagage, P. -O. and Ray, T. and Vandenbussche, B. and Waelkens, C. and Navarro-Carrera, R. and Kokorev, V.},
	month = aug,
	year = {2023},
	note = {Publisher: IOP
ADS Bibcode: 2023ApJ...952..143R},
	pages = {143},
}

@article{endsley_star-forming_2024,
	title = {The star-forming and ionizing properties of dwarf z 6-9 galaxies in {JADES}: insights on bursty star formation and ionized bubble growth},
	volume = {533},
	issn = {0035-8711},
	shorttitle = {The star-forming and ionizing properties of dwarf z 6-9 galaxies in {JADES}},
	url = {https://ui.adsabs.harvard.edu/abs/2024MNRAS.533.1111E},
	doi = {10.1093/mnras/stae1857},
	urldate = {2024-09-14},
	journal = {Monthly Notices of the Royal Astronomical Society},
	author = {Endsley, Ryan and Stark, Daniel P. and Whitler, Lily and Topping, Michael W. and Johnson, Benjamin D. and Robertson, Brant and Tacchella, Sandro and Alberts, Stacey and Baker, William M. and Bhatawdekar, Rachana and Boyett, Kristan and Bunker, Andrew J. and Cameron, Alex J. and Carniani, Stefano and Charlot, Stephane and Chen, Zuyi and Chevallard, Jacopo and Curtis-Lake, Emma and Danhaive, A. Lola and Egami, Eiichi and Eisenstein, Daniel J. and Hainline, Kevin and Helton, Jakob M. and Ji, Zhiyuan and Looser, Tobias J. and Maiolino, Roberto and Nelson, Erica and Puskás, Dávid and Rieke, George and Rieke, Marcia and Rix, Hans-Walter and Sandles, Lester and Saxena, Aayush and Simmonds, Charlotte and Smit, Renske and Sun, Fengwu and Williams, Christina C. and Willmer, Christopher N. A. and Willott, Chris and Witstok, Joris},
	month = sep,
	year = {2024},
	note = {Publisher: OUP
ADS Bibcode: 2024MNRAS.533.1111E},
	pages = {1111--1142},
}

@article{matthee_little_2024,
	title = {Little {Red} {Dots}: {An} {Abundant} {Population} of {Faint} {Active} {Galactic} {Nuclei} at z ∼ 5 {Revealed} by the {EIGER} and {FRESCO} {JWST} {Surveys}},
	volume = {963},
	issn = {0004-637X},
	shorttitle = {Little {Red} {Dots}},
	url = {https://ui.adsabs.harvard.edu/abs/2024ApJ...963..129M},
	doi = {10.3847/1538-4357/ad2345},
	urldate = {2024-09-14},
	journal = {The Astrophysical Journal},
	author = {Matthee, Jorryt and Naidu, Rohan P. and Brammer, Gabriel and Chisholm, John and Eilers, Anna-Christina and Goulding, Andy and Greene, Jenny and Kashino, Daichi and Labbe, Ivo and Lilly, Simon J. and Mackenzie, Ruari and Oesch, Pascal A. and Weibel, Andrea and Wuyts, Stijn and Xiao, Mengyuan and Bordoloi, Rongmon and Bouwens, Rychard and van Dokkum, Pieter and Illingworth, Garth and Kramarenko, Ivan and Maseda, Michael V. and Mason, Charlotte and Meyer, Romain A. and Nelson, Erica J. and Reddy, Naveen A. and Shivaei, Irene and Simcoe, Robert A. and Yue, Minghao},
	month = mar,
	year = {2024},
	note = {},
	pages = {129},
}

@article{maiolino_small_2024,
	title = {A small and vigorous black hole in the early {Universe}},
	volume = {627},
	issn = {0028-0836},
	url = {https://ui.adsabs.harvard.edu/abs/2024Natur.627...59M},
	doi = {10.1038/s41586-024-07052-5},
	urldate = {2024-09-14},
	journal = {Nature},
	author = {Maiolino, Roberto and Scholtz, Jan and Witstok, Joris and Carniani, Stefano and D'Eugenio, Francesco and de Graaff, Anna and Übler, Hannah and Tacchella, Sandro and Curtis-Lake, Emma and Arribas, Santiago and Bunker, Andrew and Charlot, Stéphane and Chevallard, Jacopo and Curti, Mirko and Looser, Tobias J. and Maseda, Michael V. and Rawle, Timothy D. and Rodríguez del Pino, Bruno and Willott, Chris J. and Egami, Eiichi and Eisenstein, Daniel J. and Hainline, Kevin N. and Robertson, Brant and Williams, Christina C. and Willmer, Christopher N. A. and Baker, William M. and Boyett, Kristan and DeCoursey, Christa and Fabian, Andrew C. and Helton, Jakob M. and Ji, Zhiyuan and Jones, Gareth C. and Kumari, Nimisha and Laporte, Nicolas and Nelson, Erica J. and Perna, Michele and Sandles, Lester and Shivaei, Irene and Sun, Fengwu},
	month = mar,
	year = {2024},
	note = {},
	pages = {59--63},
}

@ARTICLE{PMontero2013,
       author = {{P{\'e}rez-Montero}, E. and {Contini}, T. and {Lamareille}, F. and {Maier}, C. and {Carollo}, C.~M. and {Kneib}, J.-P. and {Le F{\`e}vre}, O. and {Lilly}, S. and {Mainieri}, V. and {Renzini}, A. and {Scodeggio}, M. and {Zamorani}, G. and {Bardelli}, S. and {Bolzonella}, M. and {Bongiorno}, A. and {Caputi}, K. and {Cucciati}, O. and {de la Torre}, S. and {de Ravel}, L. and {Franzetti}, P. and {Garilli}, B. and {Iovino}, A. and {Kampczyk}, P. and {Knobel}, C. and {Kova{\v{c}}}, K. and {Le Borgne}, J.-F. and {Le Brun}, V. and {Mignoli}, M. and {Pell{\`o}}, R. and {Peng}, Y. and {Presotto}, V. and {Ricciardelli}, E. and {Silverman}, J.~D. and {Tanaka}, M. and {Tasca}, L.~A.~M. and {Tresse}, L. and {Vergani}, D. and {Zucca}, E.},
        title = "{The cosmic evolution of oxygen and nitrogen abundances in star-forming galaxies over the past 10 Gyr}",
      journal = {\aap},
         year = 2013,
        month = jan,
       volume = {549},
          eid = {A25},
        pages = {A25},
          doi = {10.1051/0004-6361/201220070},
archivePrefix = {arXiv},
       eprint = {1210.0334},
 primaryClass = {astro-ph.CO},
       adsurl = {https://ui.adsabs.harvard.edu/abs/2013A&A...549A..25P}
}

@ARTICLE{Guseva2009,
       author = {{Guseva}, N.~G. and {Papaderos}, P. and {Meyer}, H.~T. and {Izotov}, Y.~I. and {Fricke}, K.~J.},
        title = "{An investigation of the luminosity-metallicity relation for a large sample of low-metallicity emission-line galaxies}",
      journal = {\aap},
         year = 2009,
        month = oct,
       volume = {505},
       number = {1},
        pages = {63-72},
          doi = {10.1051/0004-6361/200912414},
archivePrefix = {arXiv},
       eprint = {0908.2539},
 primaryClass = {astro-ph.CO},
       adsurl = {https://ui.adsabs.harvard.edu/abs/2009A&A...505...63G}
}

@ARTICLE{Amorin2014,
       author = {{Amor{\'\i}n}, R. and {Sommariva}, V. and {Castellano}, M. and {Grazian}, A. and {Tasca}, L.~A.~M. and {Fontana}, A. and {Pentericci}, L. and {Cassata}, P. and {Garilli}, B. and {Le Brun}, V. and {Le F{\`e}vre}, O. and {Maccagni}, D. and {Thomas}, R. and {Vanzella}, E. and {Zamorani}, G. and {Zucca}, E. and {Bardelli}, S. and {Capak}, P. and {Cassar{\'a}}, L.~P. and {Cimatti}, A. and {Cuby}, J.~G. and {Cucciati}, O. and {de la Torre}, S. and {Durkalec}, A. and {Giavalisco}, M. and {Hathi}, N.~P. and {Ilbert}, O. and {Lemaux}, B.~C. and {Moreau}, C. and {Paltani}, S. and {Ribeiro}, B. and {Salvato}, M. and {Schaerer}, D. and {Scodeggio}, M. and {Talia}, M. and {Taniguchi}, Y. and {Tresse}, L. and {Vergani}, D. and {Wang}, P.~W. and {Charlot}, S. and {Contini}, T. and {Fotopoulou}, S. and {L{\'o}pez-Sanjuan}, C. and {Mellier}, Y. and {Scoville}, N.},
        title = "{Discovering extremely compact and metal-poor, star-forming dwarf galaxies out to z \raisebox{-0.5ex}\textasciitilde 0.9 in the VIMOS Ultra-Deep Survey}",
      journal = {\aap},
         year = 2014,
        month = aug,
       volume = {568},
          eid = {L8},
        pages = {L8},
          doi = {10.1051/0004-6361/201423816},
archivePrefix = {arXiv},
       eprint = {1403.3692},
 primaryClass = {astro-ph.GA},
       adsurl = {https://ui.adsabs.harvard.edu/abs/2014A&A...568L...8A}
}

@ARTICLE{Leboutellier2013,
       author = {{Lebouteiller}, V. and {Heap}, S. and {Hubeny}, I. and {Kunth}, D.},
        title = "{Chemical enrichment and physical conditions in I Zw 18}",
      journal = {\aap},
         year = 2013,
        month = may,
       volume = {553},
          eid = {A16},
        pages = {A16},
          doi = {10.1051/0004-6361/201220948},
archivePrefix = {arXiv},
       eprint = {1302.4746},
 primaryClass = {astro-ph.CO},
       adsurl = {https://ui.adsabs.harvard.edu/abs/2013A&A...553A..16L}
}

@ARTICLE{Izotov2024,
       author = {{Izotov}, Y.~I. and {Thuan}, T.~X. and {Guseva}, N.~G.},
        title = "{J1046+4047: an extremely low-metallicity dwarf star-forming galaxy with O$_{32}$ = 57}",
      journal = {\mnras},
         year = 2024,
        month = jan,
       volume = {527},
       number = {2},
        pages = {3486-3493},
          doi = {10.1093/mnras/stad3421},
archivePrefix = {arXiv},
       eprint = {2311.01799},
 primaryClass = {astro-ph.GA},
       adsurl = {https://ui.adsabs.harvard.edu/abs/2024MNRAS.527.3486I}
}

@ARTICLE{Vanzella2016,
       author = {{Vanzella}, E. and {De Barros}, S. and {Cupani}, G. and {Karman}, W. and {Gronke}, M. and {Balestra}, I. and {Coe}, D. and {Mignoli}, M. and {Brusa}, M. and {Calura}, F. and {Caminha}, G.-B. and {Caputi}, K. and {Castellano}, M. and {Christensen}, L. and {Comastri}, A. and {Cristiani}, S. and {Dijkstra}, M. and {Fontana}, A. and {Giallongo}, E. and {Giavalisco}, M. and {Gilli}, R. and {Grazian}, A. and {Grillo}, C. and {Koekemoer}, A. and {Meneghetti}, M. and {Nonino}, M. and {Pentericci}, L. and {Rosati}, P. and {Schaerer}, D. and {Verhamme}, A. and {Vignali}, C. and {Zamorani}, G.},
        title = "{High-resolution Spectroscopy of a Young, Low-metallicity Optically Thin L = 0.02L* Star-forming Galaxy at z = 3.12}",
      journal = {\apjl},
         year = 2016,
        month = apr,
       volume = {821},
       number = {2},
          eid = {L27},
        pages = {L27},
          doi = {10.3847/2041-8205/821/2/L27},
archivePrefix = {arXiv},
       eprint = {1603.01616},
 primaryClass = {astro-ph.GA},
       adsurl = {https://ui.adsabs.harvard.edu/abs/2016ApJ...821L..27V}
}

@ARTICLE{Vanzella2021,
       author = {{Vanzella}, E. and {Caminha}, G.~B. and {Rosati}, P. and {Mercurio}, A. and {Castellano}, M. and {Meneghetti}, M. and {Grillo}, C. and {Sani}, E. and {Bergamini}, P. and {Calura}, F. and {Caputi}, K. and {Cristiani}, S. and {Cupani}, G. and {Fontana}, A. and {Gilli}, R. and {Grazian}, A. and {Gronke}, M. and {Mignoli}, M. and {Nonino}, M. and {Pentericci}, L. and {Tozzi}, P. and {Treu}, T. and {Balestra}, I. and {Dijkstra}, M.},
        title = "{The MUSE Deep Lensed Field on the Hubble Frontier Field MACS J0416. Star-forming complexes at cosmological distances}",
      journal = {\aap},
         year = 2021,
        month = feb,
       volume = {646},
          eid = {A57},
        pages = {A57},
          doi = {10.1051/0004-6361/202039466},
archivePrefix = {arXiv},
       eprint = {2009.08458},
 primaryClass = {astro-ph.GA},
       adsurl = {https://ui.adsabs.harvard.edu/abs/2021A&A...646A..57V}
}

@ARTICLE{Rusta2025,
       author = {{Rusta}, Elka and {Salvadori}, Stefania and {Gelli}, Viola and {Schaerer}, Daniel and {Marconi}, Alessandro and {Koutsouridou}, Ioanna and {Carniani}, Stefano},
        title = "{Metal-polluted Population III Galaxies and How to Find Them}",
      journal = {\apjl},
         year = 2025,
        month = aug,
       volume = {989},
       number = {2},
          eid = {L32},
        pages = {L32},
          doi = {10.3847/2041-8213/adf4e3},
archivePrefix = {arXiv},
       eprint = {2506.17400},
 primaryClass = {astro-ph.GA},
       adsurl = {https://ui.adsabs.harvard.edu/abs/2025ApJ...989L..32R}
}

@ARTICLE{Sarmento2017,
       author = {{Sarmento}, Richard and {Scannapieco}, Evan and {Pan}, Liubin},
        title = "{Following the Cosmic Evolution of Pristine Gas. I. Implications for Milky Way Halo Stars}",
      journal = {\apj},
         year = 2017,
        month = jan,
       volume = {834},
       number = {1},
          eid = {23},
        pages = {23},
          doi = {10.3847/1538-4357/834/1/23},
archivePrefix = {arXiv},
       eprint = {1611.00025},
 primaryClass = {astro-ph.GA},
       adsurl = {https://ui.adsabs.harvard.edu/abs/2017ApJ...834...23S}
}

@ARTICLE{Schaerer2022,
       author = {{Schaerer}, D. and {Marques-Chaves}, R. and {Barrufet}, L. and {Oesch}, P. and {Izotov}, Y.~I. and {Naidu}, R. and {Guseva}, N.~G. and {Brammer}, G.},
        title = "{First look with JWST spectroscopy: Resemblance among z {\ensuremath{\sim}} 8 galaxies and local analogs}",
      journal = {\aap},
         year = 2022,
        month = sep,
       volume = {665},
          eid = {L4},
        pages = {L4},
          doi = {10.1051/0004-6361/202244556},
archivePrefix = {arXiv},
       eprint = {2207.10034},
 primaryClass = {astro-ph.GA},
       adsurl = {https://ui.adsabs.harvard.edu/abs/2022A&A...665L...4S}
}

@ARTICLE{Castellano2024,
       author = {{Castellano}, Marco and {Napolitano}, Lorenzo and {Fontana}, Adriano and {Roberts-Borsani}, Guido and {Treu}, Tommaso and {Vanzella}, Eros and {Zavala}, Jorge A. and {Arrabal Haro}, Pablo and {Calabr{\`o}}, Antonello and {Llerena}, Mario and {Mascia}, Sara and {Merlin}, Emiliano and {Paris}, Diego and {Pentericci}, Laura and {Santini}, Paola and {Bakx}, Tom J.~L.~C. and {Bergamini}, Pietro and {Cupani}, Guido and {Dickinson}, Mark and {Filippenko}, Alexei V. and {Glazebrook}, Karl and {Grillo}, Claudio and {Kelly}, Patrick L. and {Malkan}, Matthew A. and {Mason}, Charlotte A. and {Morishita}, Takahiro and {Nanayakkara}, Themiya and {Rosati}, Piero and {Sani}, Eleonora and {Wang}, Xin and {Yoon}, Ilsang},
        title = "{JWST NIRSpec Spectroscopy of the Remarkable Bright Galaxy GHZ2/GLASS-z12 at Redshift 12.34}",
      journal = {\apj},
         year = 2024,
        month = sep,
       volume = {972},
       number = {2},
          eid = {143},
        pages = {143},
          doi = {10.3847/1538-4357/ad5f88},
archivePrefix = {arXiv},
       eprint = {2403.10238},
 primaryClass = {astro-ph.GA},
       adsurl = {https://ui.adsabs.harvard.edu/abs/2024ApJ...972..143C}
}

@ARTICLE{Maiolino2025,
       author = {{Maiolino}, Roberto and {Uebler}, Hannah and {D'Eugenio}, Francesco and {Scholtz}, Jan and {Juodzbalis}, Ignas and {Ji}, Xihan and {Perna}, Michele and {Bromm}, Volker and {Dayal}, Pratika and {Koudmani}, Sophie and {Liu}, Boyuan and {Schneider}, Raffaella and {Sijacki}, Debora and {Valiante}, Rosa and {Trinca}, Alessandro and {Zhang}, Saiyang and {Volonteri}, Marta and {Inayoshi}, Kohei and {Carniani}, Stefano and {Nakajima}, Kimihiko and {Isobe}, Yuki and {Witstok}, Joris and {Jones}, Gareth C. and {Tacchella}, Sandro and {Arribas}, Santiago and {Bunker}, Andrew and {Cataldi}, Elisa and {Charlot}, Stephane and {Cresci}, Giovanni and {Curti}, Mirko and {Fabian}, Andrew C. and {Katz}, Harley and {Kumari}, Nimisha and {Laporte}, Nicolas and {Mazzolari}, Giovanni and {Robertson}, Brant and {Sun}, Fengwu and {Rodriguez Del Pino}, Bruno and {Venturi}, Giacomo},
        title = "{A black hole in a near-pristine galaxy 700 million years after the Big Bang}",
      journal = {arXiv e-prints},
         year = 2025,
        month = may,
          eid = {arXiv:2505.22567},
        pages = {arXiv:2505.22567},
          doi = {10.48550/arXiv.2505.22567},
archivePrefix = {arXiv},
       eprint = {2505.22567},
 primaryClass = {astro-ph.GA},
       adsurl = {https://ui.adsabs.harvard.edu/abs/2025arXiv250522567M}
}

@ARTICLE{Morishita2025,
       author = {{Morishita}, Takahiro and {Liu}, Zhaoran and {Stiavelli}, Massimo and {Treu}, Tommaso and {Bergamini}, Pietro and {Zhang}, Yechi},
        title = "{Pristine Massive Star Formation Caught at the Break of Cosmic Dawn}",
      journal = {arXiv e-prints},
         year = 2025,
        month = jul,
          eid = {arXiv:2507.10521},
        pages = {arXiv:2507.10521},
          doi = {10.48550/arXiv.2507.10521},
archivePrefix = {arXiv},
       eprint = {2507.10521},
 primaryClass = {astro-ph.CO},
       adsurl = {https://ui.adsabs.harvard.edu/abs/2025arXiv250710521M}
}

@ARTICLE{Vanzella2023,
       author = {{Vanzella}, E. and {Loiacono}, F. and {Bergamini}, P. and {Me{\v{s}}tri{\'c}}, U. and {Castellano}, M. and {Rosati}, P. and {Meneghetti}, M. and {Grillo}, C. and {Calura}, F. and {Mignoli}, M. and {Brada{\v{c}}}, M. and {Adamo}, A. and {Rihtar{\v{s}}i{\v{c}}}, G. and {Dickinson}, M. and {Gronke}, M. and {Zanella}, A. and {Annibali}, F. and {Willott}, C. and {Messa}, M. and {Sani}, E. and {Acebron}, A. and {Bolamperti}, A. and {Comastri}, A. and {Gilli}, R. and {Caputi}, K.~I. and {Ricotti}, M. and {Gruppioni}, C. and {Ravindranath}, S. and {Mercurio}, A. and {Strait}, V. and {Martis}, N. and {Pascale}, R. and {Caminha}, G.~B. and {Annunziatella}, M. and {Nonino}, M.},
        title = "{An extremely metal-poor star complex in the reionization era: Approaching Population III stars with JWST}",
      journal = {\aap},
         year = 2023,
        month = oct,
       volume = {678},
          eid = {A173},
        pages = {A173},
          doi = {10.1051/0004-6361/202346981},
archivePrefix = {arXiv},
       eprint = {2305.14413},
 primaryClass = {astro-ph.GA},
       adsurl = {https://ui.adsabs.harvard.edu/abs/2023A&A...678A.173V}
}

@ARTICLE{Taylor2025,
       author = {{Taylor}, Anthony J. and {Kokorev}, Vasily and {Kocevski}, Dale D. and {Akins}, Hollis B. and {Cullen}, Fergus and {Dickinson}, Mark and {Finkelstein}, Steven L. and {Arrabal Haro}, Pablo and {Bromm}, Volker and {Giavalisco}, Mauro and {Inayoshi}, Kohei and {Juneau}, St{\'e}phanie and {Leung}, Gene C.~K. and {P{\'e}rez-Gonz{\'a}lez}, Pablo G. and {Somerville}, Rachel S. and {Trump}, Jonathan R. and {Amor{\'\i}n}, Ricardo O. and {Barro}, Guillermo and {Burgarella}, Denis and {Brooks}, Madisyn and {Carnall}, Adam C. and {Casey}, Caitlin M. and {Cheng}, Yingjie and {Chisholm}, John and {Chworowsky}, Katherine and {Davis}, Kelcey and {Donnan}, Callum T. and {Dunlop}, James S. and {Ellis}, Richard S. and {Fern{\'a}ndez}, Vital and {Fujimoto}, Seiji and {Grogin}, Norman A. and {Gupta}, Ansh R. and {Hathi}, Nimish P. and {Jung}, Intae and {Hirschmann}, Michaela and {Kartaltepe}, Jeyhan S. and {Koekemoer}, Anton M. and {Larson}, Rebecca L. and {Leung}, Ho-Hin and {Llerena}, Mario and {Lucas}, Ray A. and {McLeod}, Derek J. and {McLure}, Ross and {Napolitano}, Lorenzo and {Papovich}, Casey and {Stanton}, Thomas M. and {Tripodi}, Roberta and {Wang}, Xin and {Wilkins}, Stephen M. and {Yung}, L.~Y. Aaron and {Zavala}, Jorge A.},
        title = "{CAPERS-LRD-z9: A Gas-enshrouded Little Red Dot Hosting a Broad-line Active Galactic Nucleus at z = 9.288}",
      journal = {\apjl},
         year = 2025,
        month = aug,
       volume = {989},
       number = {1},
          eid = {L7},
        pages = {L7},
          doi = {10.3847/2041-8213/ade789},
archivePrefix = {arXiv},
       eprint = {2505.04609},
 primaryClass = {astro-ph.GA},
       adsurl = {https://ui.adsabs.harvard.edu/abs/2025ApJ...989L...7T}
}

@ARTICLE{Lotz2017,
       author = {{Lotz}, J.~M. and {Koekemoer}, A. and {Coe}, D. and {Grogin}, N. and {Capak}, P. and {Mack}, J. and {Anderson}, J. and {Avila}, R. and {Barker}, E.~A. and {Borncamp}, D. and {Brammer}, G. and {Durbin}, M. and {Gunning}, H. and {Hilbert}, B. and {Jenkner}, H. and {Khandrika}, H. and {Levay}, Z. and {Lucas}, R.~A. and {MacKenty}, J. and {Ogaz}, S. and {Porterfield}, B. and {Reid}, N. and {Robberto}, M. and {Royle}, P. and {Smith}, L.~J. and {Storrie-Lombardi}, L.~J. and {Sunnquist}, B. and {Surace}, J. and {Taylor}, D.~C. and {Williams}, R. and {Bullock}, J. and {Dickinson}, M. and {Finkelstein}, S. and {Natarajan}, P. and {Richard}, J. and {Robertson}, B. and {Tumlinson}, J. and {Zitrin}, A. and {Flanagan}, K. and {Sembach}, K. and {Soifer}, B.~T. and {Mountain}, M.},
        title = "{The Frontier Fields: Survey Design and Initial Results}",
      journal = {\apj},
         year = 2017,
        month = mar,
       volume = {837},
       number = {1},
          eid = {97},
        pages = {97},
          doi = {10.3847/1538-4357/837/1/97},
archivePrefix = {arXiv},
       eprint = {1605.06567},
 primaryClass = {astro-ph.GA},
       adsurl = {https://ui.adsabs.harvard.edu/abs/2017ApJ...837...97L}
}

@ARTICLE{Richard2021,
       author = {{Richard}, Johan and {Claeyssens}, Ad{\'e}la{\"\i}de and {Lagattuta}, David and {Guaita}, Lucia and {Bauer}, Franz Erik and {Pello}, Roser and {Carton}, David and {Bacon}, Roland and {Soucail}, Genevi{\`e}ve and {Lyon}, Gonzalo Prieto and {Kneib}, Jean-Paul and {Mahler}, Guillaume and {Cl{\'e}ment}, Benjamin and {Mercier}, Wilfried and {Variu}, Andrei and {Tamone}, Am{\'e}lie and {Ebeling}, Harald and {Schmidt}, Kasper B. and {Nanayakkara}, Themiya and {Maseda}, Michael and {Weilbacher}, Peter M. and {Bouch{\'e}}, Nicolas and {Bouwens}, Rychard J. and {Wisotzki}, Lutz and {de la Vieuville}, Geoffroy and {Martinez}, Johany and {Patr{\'\i}cio}, Vera},
        title = "{An atlas of MUSE observations towards twelve massive lensing clusters}",
      journal = {\aap},
         year = 2021,
        month = feb,
       volume = {646},
          eid = {A83},
        pages = {A83},
          doi = {10.1051/0004-6361/202039462},
archivePrefix = {arXiv},
       eprint = {2009.09784},
 primaryClass = {astro-ph.GA},
       adsurl = {https://ui.adsabs.harvard.edu/abs/2021A&A...646A..83R}
}

@ARTICLE{Sarrouh2026,
       author = {{Sarrouh}, Ghassan T.~E. and {Asada}, Yoshihisa and {Martis}, Nicholas S. and {Willott}, Chris J. and {Iyer}, Kartheik G. and {Noirot}, Ga{\"e}l and {Muzzin}, Adam and {Sawicki}, Marcin and {Brammer}, Gabriel and {Desprez}, Guillaume and {Rihtar{\v{s}}i{\v{c}}}, Gregor and {Zabl}, Johannes and {Abraham}, Roberto and {Brada{\v{c}}}, Maru{\v{s}}a and {Doyon}, Ren{\'e} and {Antwi-Danso}, Jacqueline and {Berek}, Samantha and {Brown}, Westley and {Estrada-Carpenter}, Vince and {Favaro}, Jeremy and {Felicioni}, Giordano and {Forrest}, Ben and {Gaspar}, Gaia and {Gould}, Katriona M.~L. and {Gledhill}, Rachel and {Harshan}, Anishya and {Jahan}, Nusrath and {Jagga}, Naadiyah and {Jude{\v{z}}}, Jon and {Marchesini}, Danilo and {Markov}, Vladan and {Matharu}, Jasleen and {MacFarland}, Shannon and {Merchant}, Maya and {M{\'e}rida}, Rosa M. and {Mowla}, Lamiya and {Myers}, Katherine and {Omori}, Kiyoaki C. and {Pacifici}, Camilla and {Ravindranath}, Swara and {Robbins}, Luke and {Strait}, Victoria and {Sok}, Visal and {Tan}, Vivian Yun Yan and {Tripodi}, Roberta and {Wilson}, Gillian and {Withers}, Sunna},
        title = "{CANUCS/Technicolor Data Release 1: Imaging, Photometry, Slit Spectroscopy, and Stellar Population Parameters}",
      journal = {\apjs},
         year = 2026,
        month = jan,
       volume = {282},
       number = {1},
          eid = {3},
        pages = {3},
          doi = {10.3847/1538-4365/ae1611},
archivePrefix = {arXiv},
       eprint = {2506.21685},
 primaryClass = {astro-ph.GA},
       adsurl = {https://ui.adsabs.harvard.edu/abs/2026ApJS..282....3S}
}

@ARTICLE{Gledhill2024,
       author = {{Gledhill}, Rachel and {Strait}, Victoria and {Desprez}, Guillaume and {Rihtar{\v{s}}i{\v{c}}}, Gregor and {Brada{\v{c}}}, Maru{\v{s}}a and {Brammer}, Gabriel and {Willott}, Chris J. and {Martis}, Nicholas and {Sawicki}, Marcin and {Noirot}, Ga{\"e}l and {Sarrouh}, Ghassan T.~E. and {Muzzin}, Adam},
        title = "{CANUCS: An Updated Mass and Magnification Model of A370 with JWST}",
      journal = {\apj},
         year = 2024,
        month = oct,
       volume = {973},
       number = {2},
          eid = {77},
        pages = {77},
          doi = {10.3847/1538-4357/ad684a},
archivePrefix = {arXiv},
       eprint = {2403.07062},
 primaryClass = {astro-ph.GA},
       adsurl = {https://ui.adsabs.harvard.edu/abs/2024ApJ...973...77G}
}

@ARTICLE{Cooper2025,
       author = {{Cooper}, Ryan A. and {Caputi}, Karina I. and {Iani}, Edoardo and {Rinaldi}, Pierluigi and {Desprez}, Guillaume and {Navarro-Carrera}, Rafael},
        title = "{High-velocity Outflows in [O III] Emitters at 2.5 < z < 9 from JWST NIRSpec Medium-resolution Spectroscopy}",
      journal = {\apj},
         year = 2025,
        month = nov,
       volume = {994},
       number = {1},
          eid = {102},
        pages = {102},
          doi = {10.3847/1538-4357/ae0580},
archivePrefix = {arXiv},
       eprint = {2502.18310},
 primaryClass = {astro-ph.GA},
       adsurl = {https://ui.adsabs.harvard.edu/abs/2025ApJ...994..102C}
}
\bibliographystyle{aasjournal}

\end{document}